\documentclass[aps,prd,preprint,eqsecnum,nofootinbib,groupedaddress,showpacs,floatfix]{revtex4}
\usepackage[dvips]{graphicx}
\usepackage{multirow}
\font\ttiny=cmr6 at 3.5pt

\newcommand{\BlackHat}{{\sc BlackHat}}
\newcommand{\SHERPA}{{\sc SHERPA}}
\newcommand{\AMEGIC}{{\sc AMEGIC++\/}}
\newcommand{\SISCone}{{\sc SISCone}}
\newcommand{\MCFM}{{\sc MCFM}}
\newcommand{\Vegas}{{\sc Vegas}}

\newif\ifdraft
\drafttrue
\newif\ifpreprint
\preprinttrue

\def\fig#1{fig.~{\ref{#1}}}
\def\Fig#1{Fig.~{\ref{#1}}}
\def\figs#1#2{figs.~{\ref{#1}} and {\ref{#2}}}

\def\sect#1{section~{\ref{#1}}}
\def\eqn#1{eq.~(\ref{#1})}

\def\eqns#1#2{eqs.~(\ref{#1}) and~(\ref{#2})}

\def\tab#1{table~{\ref{#1}}}
\def\Tab#1{Table~{\ref{#1}}}
\def\qb{\bar q}
\def\pb{\bar p}

\def\Qb{\bar Q}
\def\nub{\bar \nu}
\def\e{\epsilon}
\def\bom#1{{\mbox{\boldmath $#1$}}}
\def\ETslash{{\s E}_T}
\def\HTpartonic{{\hat H}_T}
\def\ETW{E_{T}^{W}}
\def\Wj{$W\,\!+\,1$}
\def\Wjj{$W\,\!+\,2$}
\def\Wjjj{$W\,\!+\,3$}
\def\Wjjjj{$W\,\!+\,4$}
\def\Wjx{$W\,\!+\,1,2$}
\def\Wjjx{$W\,\!+\,2,3$}

\def\Wjn{$W\,\!+\,n$}

\def\Wmjjj{$W^-\,\!+\,3$}

\def\Wmjjja{$W^-\,\!+\,1,2,3$}
\def\Wmjn{$W^-\,\!+\,n$}

\def\Wpjj{$W^+\,\!+\,2$}
\def\Wpjjj{$W^+\,\!+\,3$}

\def\Wpjjx{$W^+\,\!+\,2,3$}
\def\Wpjjja{$W^+\,\!+\,1,2,3$}
\def\Wpjn{$W^+\,\!+\,n$}

\def\Zjj{$Z\,\!+\,2$}
\def\Zjjj{$Z\,\!+\,3$}
\def\Zjjjj{$Z\,\!+\,4$}

\def\Zjjjj{$Z\,\!+\,4$}
\def\ETsl{{\s E}_T}
\def\jets{\,{\rm jets}\,}
\def\jet{{\rm jet}}
\def\eps{\epsilon}
\def\nn{\nonumber}
\def\MLO{d\sigma^{(0)}}
\def\MNLO{d\sigma_V^{(1)}}
\def\hatMNLO{\widehat{d\sigma}_V^{(1)}}
\def\extraskip{\vskip 15mm}

\newbox\charbox
\newbox\slabox
\def\s#1{{      
        \setbox\charbox=\hbox{$#1$}
        \setbox\slabox=\hbox{$/$}
        \dimen\charbox=\ht\slabox
        \advance\dimen\charbox by -\dp\slabox
        \advance\dimen\charbox by -\ht\charbox
        \advance\dimen\charbox by \dp\charbox
        \divide\dimen\charbox by 2
        \raise-\dimen\charbox\hbox to \wd\charbox{\hss/\hss}
        \llap{$#1$}
}}

\def\nuornub{{}^{\raise1.3pt\hbox{\ttiny(}}\hskip -0.2pt\overline{\kern 
-0.5pt \nu \kern -0.4pt}\hskip 0.3pt{}^{\raise1.3pt\hbox{\ttiny)}}}

\begin{document}

\hbox{
UCLA/09/TEP/53 $\null\hskip 2.7cm \null$
MIT-CTP 4047 $\hskip 2.7cm \null$
Saclay--IPhT--T09/078}

\hbox{
IPPP/09/46 $\hskip 10.3cm \null$
SLAC--PUB--13680
}

\title{Next-to-Leading Order QCD Predictions for $\bom{W}$\,+\,3-Jet
Distributions at Hadron Colliders}

\author{C.~F.~Berger${}^{a}$, Z.~Bern${}^b$,
L.~J.~Dixon${}^c$, F.~Febres Cordero${}^b$, D.~Forde${}^{c}$,
T.~Gleisberg${}^c$,  H. Ita${}^b$,
D.~A.~Kosower${}^d$ and D.~Ma\^{\i}tre${}^e$}%
\affiliation{\centerline{${}^a${Center for Theoretical
Physics, Massachusetts Institute of Technology,
      Cambridge, MA 02139, USA}} \\
\centerline{${}^b$Department of Physics and Astronomy, UCLA, Los Angeles, CA
90095-1547, USA} \\
\centerline{${}^c$SLAC National Accelerator Laboratory, Stanford University,
             Stanford, CA 94309, USA} \\
\centerline{${}^d$Institut de Physique Th\'eorique, CEA--Saclay,
          F--91191 Gif-sur-Yvette cedex, France}\\
\centerline{${}^e$Department of Physics, University of Durham,
          DH1 3LE, UK}
}

\begin{abstract}

We present next-to-leading order QCD predictions for a variety of
distributions in \Wjjj-jet production at both the Tevatron and the
Large Hadron Collider.  We include all subprocesses and incorporate
the decay of the $W$ boson into leptons.  Our results are in excellent
agreement with existing Tevatron data and provide the first
quantitatively precise next-to-leading order predictions for the LHC.
We include all terms in an expansion in the number of colors,
confirming that the specific leading-color approximation used in our
previous study is accurate to within three percent.  The dependence
of the cross section on renormalization and factorization scales is
reduced significantly with respect to a leading-order calculation.  We
study different dynamical scale choices, and find that the total
transverse energy is significantly better than choices
used in previous phenomenological studies.  We compute the one-loop matrix
elements using on-shell methods, as numerically implemented in the
\BlackHat{} code.  The remaining parts of the calculation, including
generation of the real-emission contributions and integration over
phase space, are handled by the \SHERPA{} package. 
\end{abstract}

\pacs{12.38.-t, 12.38.Bx, 13.87.-a, 14.70.-e, 14.70.Fm, 11.15.-q,
11.15.Bt, 11.55.-m \hspace{1cm}}

\maketitle

\section{Introduction}

The upcoming start of physics runs at the Large Hadron Collider (LHC)
has added impetus to the long-standing quest to improve theoretical
control over Standard-Model backgrounds to new physics searches
at hadron colliders.  Some
backgrounds can be understood without much theoretical input.
For example, a light Higgs boson decaying into two photons produces
a narrow bump in the di-photon invariant mass, from which
the large but smooth QCD background can be subtracted experimentally
using sideband information.  However, for many searches the signals
are excesses in broader distributions of jets, along with missing
energy and charged leptons or photons; such searches require a much more
detailed theoretical understanding of the QCD backgrounds.
A classic example is the production of a Higgs boson in association
with a $W$ boson at the Tevatron, with the Higgs decaying to a
$b\bar{b}$ pair, and the $W$ decaying to a charged lepton and a
neutrino.  The peak in the $b\bar{b}$ invariant mass is much broader
than in the di-photon one; therefore variations in the backgrounds,
including QCD production of $Wb\bar{b}$, across the signal region are more
difficult to assess.

In this paper, we focus on a related important class of
backgrounds, production of multiple (untagged)
jets in association with a $W$ boson.
Such events, with a leptonically decaying $W$, form a
background to supersymmetry searches at the LHC that require a lepton,
missing transverse energy and jets~\cite{MET}. 
If the lepton is missed,
they also contribute to the background for similar searches not
requiring a lepton.  The rate of events
containing a $W$ along with multiple jets can be used
to calibrate the corresponding rate for $Z$ production with multiple jets,
which form another important source of missing transverse energy
when the $Z$ decays to a pair of neutrinos.
Analysis of $W$ plus multi-jet production will also assist in
separating these events from the production of top-quark pairs,
so that more detailed studies of the latter can be performed.

The first step toward a theoretical understanding of QCD backgrounds
is the evaluation of the cross section at leading order (LO) in the
strong coupling $\alpha_S$. Our particular focus is on high jet
multiplicity in association with vector boson production. Many
computer codes~\cite{LOPrograms,HELAC,Amegic} are available to
generate predictions at leading order.  Some of the codes incorporate
higher-multiplicity leading-order matrix elements into parton
showering programs~\cite{PYTHIAetc,Sherpa}, using matching (or
merging) procedures~\cite{Matching,MLMSMPR}.  LO predictions suffer
from large renormalization- and factorization-scale dependence,
growing with increasing jet multiplicity, and already up to a factor
of two in the processes we shall study.  Next-to-leading order (NLO)
corrections are necessary to obtain quantitatively reliable
predictions.  They typically display a greatly reduced scale
dependence~\cite{LesHouches}.  Fixed-order results at NLO can also be
matched to parton showers.  This has been done for a growing list of
processes within the {\sc MC@NLO} program and the {\sc POWHEG}
method~\cite{MCNLO}.  It would be desirable to extend this matching to
higher-multiplicity processes such as those we discuss in the present
paper.

The production of massive vector bosons in association with jets at
hadron colliders has been the subject of theoretical studies for over
three decades.  Early studies were of large transverse-momentum
muon-pair production at leading order~\cite{EarlyVplus1}, followed by
the leading-order matrix elements for \Wjj-jet
production~\cite{Wplus2ME} and corresponding phenomenological
studies~\cite{EarlyWplus2,EarlyWplus2MP}.  The early leading-order
studies were followed by NLO predictions for vector boson production
in association with a single jet~\cite{Vplus1NLO, Vplus1NLOAR}.
Leading-order results for vector-boson production accompanied by three
or four jets appeared soon thereafter~\cite{Vecbos}.  These
theoretical studies played an important role in the discovery of the
top quark~\cite{TopQuarkDiscovery}. Modern matrix element
generators~\cite{LOPrograms,HELAC,Amegic} allow for even larger
numbers of final-state jets at LO.  The one-loop matrix elements for
\Wjj-jet and \Zjj-jet production were determined~\cite{Zqqgg} via the
unitarity method~\cite{UnitarityMethod} (see also
ref.~\cite{OtherZpppp}), and incorporated into the parton-level
\MCFM~\cite{MCFM} code.

Studies of $W$ production in association with heavy quarks have
also been performed.  NLO results for hadronic production of
a $W$ and a charm quark first appeared in ref.~\cite{WcNLO}.
More recently, NLO results have been presented for
$W\,\!+\,b\,+\,$jet production~\cite{WbjNLO},
as well as for $Wb\bar b$ production with full
$b$ quark mass effects~\cite{Wplus2MassiveNLO}.
The last two computations were combined to produce a full
description of $W$ production in association with
a single $b$-jet in ref.~\cite{WbNLO}.

NLO studies of $W$ production in association with more jets have long
been desirable.  However, a bottleneck to these studies was posed by
one-loop amplitudes involving six or more partons~\cite{LesHouches}.
On-shell methods~\cite{OnShellReview},
which exploit unitarity and recursion relations,
have successfully broken this bottleneck, by avoiding
gauge-noninvariant intermediate steps, and reducing the problem
to much smaller elements analogous to tree-level amplitudes.
Approaches based on Feynman diagrams have
also led to new results with six external partons, exemplified by the
NLO cross section for producing $t \bar t b \bar b$
at hadron colliders~\cite{BDDP}.
We expect that on-shell methods will be particularly advantageous
for processes involving many external gluons, which often dominate
multi-jet final states.  Various
results~\cite{CutTools,BlackHatI,GZ,OtherLargeN,HPP}
already indicate the suitability of these methods for a
general-purpose numerical approach to high-multiplicity
one-loop amplitudes.

We recently presented the first NLO results for \Wjjj-jet production
including all subprocesses~\cite{PRLW3BH}, using one-loop amplitudes
obtained by on-shell methods.  This study used a specific type of
leading-color approximation designed to have small corrections---under
3 percent, as verified in \Wjx-jet production---while reducing the
required computer time.  The study was performed for the Tevatron,
with the same cuts employed by the CDF collaboration in their
measurement of \Wjn-jet production~\cite{WCDF}.  The NLO corrections
show a much-reduced dependence on the renormalization and
factorization scales, and excellent agreement with the CDF data for
the distribution in the transverse energy $E_T$ of the third-most
energetic jet.

In the present paper, we continue our study of \Wjjj-jet production.
We present results for \Wjjj-jet production at the LHC as well as at
the Tevatron.  As before, we include all subprocesses and take all
quarks to be massless.  (We do not include top-quark contributions,
but expect them to be very small for the distributions we shall
present.)  We extend the previous results by including specific
virtual contributions that are subleading in the number of colors,
which we had previously neglected.  We shall demonstrate that, as
expected, these subleading-color corrections to cross sections and
distributions are uniformly small, generally under three percent.  We
present three sets of distributions at the Tevatron: the
$E_T$ of the third most energetic jet, the total transverse energy
$H_T$~\cite{TotalTransverseEnergy}, and the di-jet invariant masses.
These distributions are again computed with the same cuts used by CDF.
(As discussed further in \sect{TevatronResultsSection}, we used the
infrared-safe \SISCone{} jet algorithm~\cite{SISCONE} in place of
JETCLU, the cone algorithm used by CDF.)  The code we use is
general-purpose, permitting the analysis at NLO of any infrared-safe
observable in \Wjjj-jet events.  We also present a wide variety of
distributions for the ultimate LHC energy of 14~TeV.  We find that all
the NLO cross sections and distributions display the expected
reduction in renormalization- and factorization-scale dependence
compared to the same quantities calculated at leading order.

The shapes of distributions at leading order are quite sensitive to
the functional form of the scale choice.  As expected, the change in
shape between LO and NLO distributions can be reduced by choosing
typical energy scales event-by-event for the renormalization and
factorization scales, as noted by many authors over the
years~\cite{Vplus1NLOAR,EarlyWplus2MP,DynamicalScaleChoice}.  The
vector boson transverse energy $E_{T}^W$, employed as an
event-by-event scale in previous predictions and comparisons with
data~\cite{EarlyWplus2MP,ZCDF,WCDF,PRLW3BH}, turns out to be a poor
characterization of the typical energy scale for events with large jet
transverse energies, as at the LHC.  We find that the total partonic
transverse energy is a much better choice.  Recently, similar
deficiencies in the scale choice of $E_{T}^W$ at LO have been observed
independently, and another variable, related to the invariant mass of
the final-state jets, has been proposed as a replacement~\cite{Bauer}.
Here we go further and demonstrate that for LHC energies, $E_{T}^W$ is
a poor scale choice not only at LO but also at NLO, yielding negative
differential cross sections in the tails of some distributions.  This
pathology arises from large residual logarithms induced by disparities
between momentum-transfer scales in multi-jet processes and the value
of $E_{T}^W$.

For \Wjjj-jet production, choosing the total partonic transverse
energy as the scale gives rise to shapes of distributions at LO that
are typically similar to those at NLO.  For a few \Wjjj-jet
distributions genuine NLO effects are present, and significant shape
changes remain between LO and NLO.  These differences are usually less
pronounced than in \Wjx-jet production.  In the latter cases, the LO
kinematics are more constrained, leading to significantly larger NLO
corrections.  In any event, an accurate description of the shape of
any distribution requires an NLO computation, either to confirm that
its shape is unmodified compared to LO, or to quantitatively determine
the shape change.

Ellis {\it et al.\/} have recently presented partial NLO results for
\Wjjj-jet production.  Their first calculation~\cite{EMZ} was
restricted to leading-color contributions to two-quark subprocesses,
rendering it unsuitable for phenomenological studies. Their version of
the leading-color approximation drops subleading-color terms in both
the virtual and real-emission contributions. Quite
recently~\cite{EMZ2} the same authors have added the leading-color
contributions from four-quark processes, folding in the decay of the
$W$ in the zero-width approximation.  They extended their
leading-color approximation to include $n_f$-dependent terms, and
estimated the full-color result based on the leading-order ratio of
the full-color (FC) and leading-color (LC) cross sections.  The value
of the double ratio $(\sigma^{\rm NLO,FC}/\sigma^{\rm NLO,LC})/
(\sigma^{\rm LO,FC}/\sigma^{\rm LO,LC})$ they use implicitly is quite
sensitive to the inclusion of $n_f$ terms, and as noted by the
authors, sensitive to cancellations between the two-quark and
four-quark contributions.  It is nonetheless interesting that their
estimate for the total cross section is within a few percent of both
our earlier result~\cite{PRLW3BH} and the full-color one presented in
this paper.  It would be interesting to test their estimates for
various distributions against the complete results presented here; we
leave such a comparison to future work.

Next-to-leading order cross sections are built from several
ingredients: virtual corrections, computed from the interference of
tree-level and one-loop amplitudes; real-emission corrections; and a
mechanism for isolating and integrating the infrared singularities in
the latter.  We evaluate the one-loop amplitudes needed for \Wjjj-jet
production at NLO using the \BlackHat{} library~\cite{BlackHatI}.
This library implements on-shell methods for one-loop amplitudes
numerically.  Related methods have been implemented in several other
programs~\cite{CutTools,GKM,GZ,ggttg,OtherLargeN,HPP}.  A
numerical approach to amplitudes requires attention to numerical
instabilities induced by round-off error.  We have previously verified
\BlackHat's stability for
one-loop six-, seven- and eight-gluon amplitudes~\cite{BlackHatI}, and
for leading-color amplitudes for a vector boson with up to five
partons~\cite{ICHEPBH}, using a flat distribution of phase-space points.
In the present work, we confirm the stability
of \BlackHat{}-computed matrix elements for an ensemble of
points chosen in the same way as in the actual numerical integration
of the cross section.

The real-emission corrections to the LO process arise from tree-level
amplitudes with one additional parton: an additional gluon, or a
quark--antiquark pair replacing a gluon.  To isolate and cancel the
infrared divergences that arise in the integration of these terms, we
use the Catani--Seymour dipole subtraction method~\cite{CS}, as
implemented~\cite{AutomatedAmegic} in the
program~\AMEGIC{}~\cite{Amegic}, itself part of the \SHERPA{}
framework~\cite{Sherpa}.  (We also use \AMEGIC{} for the required
tree-level matrix elements.)  Other automated implementations of the
dipole subtraction method have been presented
recently~\cite{AutomatedSubtractionOther}.

The smallness of the subleading-color corrections to the specific
leading-color approximation employed in ref.~\cite{PRLW3BH} allows us
to use a ``color-expansion sampling'' approach~\cite{NLOZ4Jets}.
In this approach the subleading-color terms, while more time-consuming
per phase-space point, are sampled with lower statistics
than the leading-color ones, and therefore do not impose an
undue burden on the computer time required.

This paper is organized as follows.  In \sect{DetailsSection} we
summarize our calculational setup, and demonstrate the numerical
stability of the one-loop matrix elements.  In
\sect{TevatronResultsSection} we present results for the Tevatron,
using the same experimental cuts as CDF.  In \sect{ScaleChoiceSection}
we discuss scale choices, showing that the choice of $W$ transverse
energy typically used for Tevatron studies can lead to significant
distortions in the shapes of distributions over the broader range of
kinematics accessible at the LHC.  We advocate instead the use of
scale choices that more accurately reflect typical energy scales in
the process, such as the total partonic transverse energy, or a fixed
fraction of it.  In \sect{LHCSection}, we present a wide variety of
distributions for the LHC.  We highlight two particular topics in
subsequent sections.  
Section~\ref{LeptonSection} examines properties of the leptons
produced by $W$ decay in \Wjjj{} jet events.  The different
pseudorapidity distributions for electrons and positrons are
presented.  Then we show the ratios, between $W^+$ and $W^-$, of the 
transverse energy distributions for both the charged leptons and neutrinos.
These two ratios have strikingly different behavior at large $E_T$ , 
presumably due to the effects of $W$ polarization.
In \sect{RapiditySection} we present results for the jet-emission
probability, as a function of the pseudorapidity separation of 
the leading two jets.  These results are relevant for searches
for the Higgs boson in vector-boson fusion production.
In \sect{ColorApproximationSection}, we discuss the specific
leading-color approximation used in our previous study, and our
approach to computing the subleading-color terms.  We give our
conclusions in \sect{ConclusionSection}.  Finally, in an appendix we
give values of squared matrix elements at a selected point
in phase space.


\extraskip
\section{Calculational Setup}
\label{DetailsSection}

At NLO, the \Wjjj-jet computation can be divided into four distinct
contributions:
\begin{itemize}
\item the leading-order contribution, requiring the tree-level
$W+5$-parton matrix elements;

\item the virtual contribution, requiring
the one-loop $W+5$-parton matrix elements (built from the
interference of one-loop and tree amplitudes);

\item the subtracted real-emission contribution,
requiring the tree-level $W+6$-parton matrix elements, an
approximation capturing their singular behavior, and integration
of the difference over the additional-emission phase space;

\item the integrated approximation (real-subtraction term),
whose infrared-singular terms
must cancel the infrared singularities in the virtual contribution.
\end{itemize}
Each of these contributions must be integrated over the
final-state phase space, after imposing appropriate cuts,
and convoluted with the initial-state parton distribution functions.

We evaluate these different contributions using a number of tools.  We
compute the virtual corrections using on-shell methods, implemented
numerically in \BlackHat{}, as outlined below.  The subtraction term
is built using Catani--Seymour dipoles~\cite{CS} as
implemented~\cite{AutomatedAmegic} in \AMEGIC{}~\cite{Amegic}.  This
matrix-element generator is part of the \SHERPA{}
package~\cite{Sherpa}.  \AMEGIC{} also provides our tree-level matrix
elements.  The phase-space integration is handled by \SHERPA, using a
multi-channel approach~\cite{MultiChannel}.
The \SHERPA{} framework makes it
simple to include various experimental cuts on phase space, and to
construct and analyze a wide variety of distributions. With this
setup, it is straightforward to make NLO predictions for any
infrared-safe physical observable.
We refer the reader to
refs.~\cite{Sherpa,Amegic,AutomatedAmegic} for descriptions of
\AMEGIC{}, \SHERPA{} and the implementation of the Catani--Seymour
dipole subtraction method.

\subsection{Subprocesses}
\label{SubprocessesSubsection}

The \Wjjj-jet process, followed by leptonic $W$ decay,
\begin{equation}
\left.   \matrix{
   \hfill{\rm Tevatron:}& p\pb\cr
   \hfill{\rm LHC:}&pp\cr} \, \right\}
  \longrightarrow
W^{\pm}+3\, \jets\rightarrow e^{\pm}\nuornub_{e}+3\, \jets\,,
\end{equation}
receives contributions from several partonic subprocesses.
At leading order, and in the virtual NLO contributions,
these subprocesses are all obtained from
\begin{eqnarray}
 && q \qb' Q \Qb g\rightarrow W^{\pm}\rightarrow e^{\pm}\, \nuornub_e \,,
\label{W3qqQQg} \\
 && q \qb' g g g\rightarrow W^{\pm}\rightarrow e^{\pm}\, \nuornub_e \,,
\label{W3qqggg}
\end{eqnarray}
by crossing three of the partons into the final state.  The $W$
couples to the $q$--$q'$ line.  We include the decay of
the vector boson ($W^{\pm}$) into a lepton pair at the amplitude
level.  The $W$ can be off shell; the lepton-pair invariant mass is
drawn from a relativistic Breit-Wigner distribution whose width is
determined by the $W$ decay rate $\Gamma_W$.  For definiteness we
present results for $W$ bosons decaying to either electrons or
positrons (plus neutrinos).  We take the leptonic decay products to be
massless; in this approximation the corresponding results for $\mu$
(and $\tau$) final states are of course identical.  Amplitudes
containing identical quarks are generated by antisymmetrizing in the
exchange of appropriate $q$ and $Q$ labels.  The light quarks,
$u,d,c,s,b$, are all treated as massless.  We do not include
contributions to the amplitudes from a real or virtual top quark; its
omission should have a very small effect on the overall result.
Except as noted below, we use the same setup for the results we
report for \Wjx-jet production.

\subsection{Color Organization of Virtual Matrix Elements}
\label{ss_ColorOrganization}

To compute the production of \Wjjj{} jets at NLO, we need the one-loop
amplitudes for the processes listed in \eqns{W3qqQQg}{W3qqggg}. 
Amplitudes in gauge theories are naturally decomposed into
a sum over permutations of terms; each term
is the product of a color factor and a color-independent kinematic
factor called a partial or color-ordered amplitude.  It is convenient
to decompose the one-loop amplitudes further,
into a set of {\it primitive\/} amplitudes~\cite{qqggg,Zqqgg}.
These are the basic gauge-invariant building blocks of the amplitude, in
which the ordering of all colored external legs is fixed,
the direction of fermion lines within the loop is fixed, and $n_f$
terms arising from fermion loops are separated out.  In \BlackHat{},
the primitive amplitudes are computed directly using the
on-shell methods reviewed in the next subsection.  The primitive
amplitudes are then combined to obtain the partial amplitudes.
The virtual contributions are assembled by summing over
interferences of the one-loop partial amplitudes with their
tree-level counterparts.

In organizing the amplitude, it is useful to keep the
numbers of colors $N_c$ and of flavors $n_f$ as parameters,
setting them to their Standard-Model values only upon evaluation.
Matrix elements, whether at tree level or at one loop, can
be organized in an expansion dictated by the $N_c\rightarrow\infty$
limit.  In this expansion, the standard ``leading-color'' contribution is the
coefficient of the leading power of $N_c$, and ``subleading-color''
refers to terms that are suppressed by at
least one power of either $1/N_c^2$, or $n_f/N_c$ from virtual quark
loops.  (The expansion in either quantity terminates at finite order,
so if all terms are kept, the result is exact in $1/N_c$.)

\begin{figure}[tbh]
\includegraphics[clip,scale=0.65]{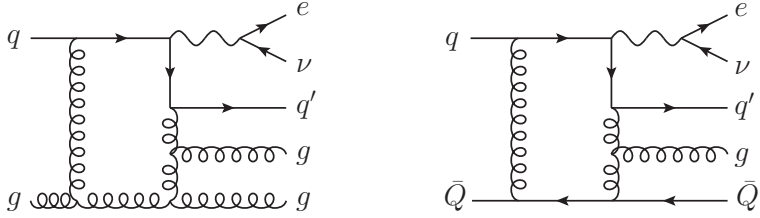}
\caption{Representative diagrams contributing at leading order in an
expansion in the number of colors to the $q g \rightarrow e \nu \, q'
\! g g$ and $q \bar Q \rightarrow e \nu \, q' \! g \bar Q$ one-loop
amplitudes.  The $e \nu$ pair couples to the quarks via a $W$ boson.
}
\label{LoopDiagramsLCFigure}
\end{figure}

\begin{figure}[tbh]
\includegraphics[clip,scale=0.59]{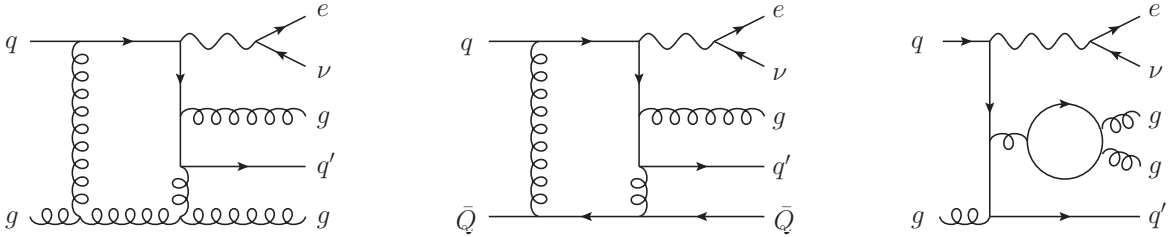}
\caption{Representative diagrams contributing only at subleading order
in an expansion in the number of colors to the $q g \rightarrow e \nu
\, q' \! g g$ and $q \bar Q \rightarrow e \nu \, q' \! g \bar Q$
one-loop amplitudes.  In such contributions, either an external gluon,
or a gluon splitting to a $\bar Q Q$ pair, is emitted from the
$q$--$q'$ line, between the $W$ boson and one of the external quarks,
$q$ or $q'$, in the cyclic ordering of the external legs.}
\label{LoopDiagramsSLCFigure}
\end{figure}

Only one primitive amplitude contributes at leading order in $1/N_c$
to each leading-color partial amplitude.
\Fig{LoopDiagramsLCFigure} shows sample ``parent'' color-ordered
Feynman diagrams for the leading-color primitive amplitudes
needed for \Wjjj-jet production.  Other diagrams contributing to
a given primitive amplitude have fewer
propagators in the loop.  They can be obtained from the diagrams
shown by moving vertices off of the loop onto trees attached to the loop,
or by using four-gluon vertices, while preserving the cyclic (color)
ordering of the external legs and the planar topology of the diagram.
In the leading-color primitive amplitudes, the $W$ boson is between
the $q$ and $q'$ external legs, with no other partons in between.

In subleading-color terms, a greater number and variety of
primitive amplitudes appear, and some primitive amplitudes
contribute to more than one subleading-color partial amplitude.
A few of the parent diagrams for subleading-color primitive amplitudes
are shown in \fig{LoopDiagramsSLCFigure}.  In such diagrams,
either another parton appears between the $W$ boson and
either $q$ or $q'$, or a gluon is emitted between $Q$ and $\bar{Q}$ in
process~(\ref{W3qqQQg}), or the diagram contains a closed fermion loop.
In the present paper, we include {\it all}
subleading-color contributions.  In \sect{ColorApproximationSection},
we discuss in greater detail how to evaluate
the full virtual cross section  efficiently, by taking advantage of the
smallness of the subleading-color contributions.


\subsection{On-Shell Methods}

The computation of one-loop partonic amplitudes has presented until
recently a bottleneck to NLO predictions for hadronic production of
four or more final-state objects (jets included).  The on-shell method
has broken this bottleneck.  This approach is based on the unitarity
method~\cite{UnitarityMethod}, including its newer refinements,
together with on-shell recursion relations~\cite{BCFW} at one
loop~\cite{Bootstrap}.  The
refinements~\cite{BCFUnitarity,OPP,EGK,Forde} rely on the use of
complex momenta, generalized unitarity and the analytic structure of
integrands, as well as subtractions to make efficient use of the known
basis of one-loop integrals.  The one-loop matrix
elements~\cite{Zqqgg} used by the \MCFM{} program~\cite{MCFM} for NLO
predictions of \Wjj-jet production were computed analytically using an
early version of this approach, and indeed served to introduce the use
of generalized unitarity~\cite{GeneralizedUnitarityOld} as an
efficient technique for loop computations.  As applied to hadron
colliders, these matrix elements have three final-state objects.
Feynman-diagram calculations have also reached into this
domain~\cite{LesHouches}.  Beyond this, improved integral reduction
techniques~\cite{IntegralReduction} have even made possible the
computation of matrix elements for four final-state
objects~\cite{BDDP,Betal} and NLO predictions using them.

Nonetheless, textbook Feynman-diagrammatic approaches suffer from a
factorial increase in complexity (or exponential if color ordered) 
and increasing degree of tensor integrals, with increasing
number of external legs.  The unitarity method for one-loop
amplitudes, in contrast, can be cast in a form with only a polynomial
increase in complexity per color-ordered helicity
configuration~\cite{Genhel,BlackHatI,GZ}.  This suggests that it will
have an increasing advantage with increasing jet multiplicity.
At fixed multiplicity, on-shell methods gain their improved
efficiency by removing {\it ab initio\/} the cancellation of
gauge-variant terms, and eliminating the need for tensor-integral (or
higher-point integral) reductions. The problem is reduced to
the computation of certain rational functions of the kinematic variables, 
to which efficient tree-like techniques can be applied.
On-shell methods have also
led to a host of analytic results, including
one-loop amplitudes in QCD with an arbitrary number of external legs,
for particular helicity assignments~\cite{Bootstrap, Genhel}.  The
reader may find reviews and further references in
refs.~\cite{OneLoopReview, OnShellReview, LesHouches}.

The \BlackHat{} library implements on-shell methods
for one-loop amplitudes numerically.  We have described the computation of
amplitudes using \BlackHat{} elsewhere~\cite{BlackHatI,ICHEPBH}.  We
limit ourselves here to an overview, along with a discussion of new
features that arise when we include subleading-color contributions
to the cross section.

Any one-loop amplitude can be written as a sum of terms containing
branch cuts in kinematic invariants, $C_n$, and terms free of
branch cuts, $R_n$,
\begin{equation}
A_n = C_n + R_n\,.
\label{CutRational}
\end{equation}
The cut part $C_n$ can in turn be written as a sum over
a basis of scalar integrals~\cite{IntegralReductions},
\begin{equation}
C_n = \sum_i d_i I_4^i + \sum_i c_i I_3^i + \sum_i b_i I_2^i \,.
\label{IntegralBasis}
\end{equation}
The scalar integrals $I_{2,3,4}^i$ --- bubbles,
triangles, and boxes --- are known
functions~\cite{IntegralsExplicit}.
They contain all the amplitude's branch cuts, packaged inside
logarithms and dilogarithms.
(Massive particles propagating in the loop would require the addition
of tadpole contributions.)
We take all external momenta to be four dimensional.
Following the spinor-helicity method~\cite{SpinorHelicity,Wplus2ME},
we can then re-express all external momenta in terms of spinors.
The coefficients of these integrals, $b_i, c_i$, and $d_i$,
as well as the rational remainder $R_n$, are then
all rational functions of
spinor variables, and more specifically of spinor products.
The problem of calculating a one-loop
amplitude then reduces to the problem of determining
these rational functions.

Generalized unitarity improves upon the original unitarity approach by
isolating smaller sets of terms, hence making use of simpler on-shell
amplitudes as basic building blocks.  Furthermore, by isolating
different integrals, it removes the need for integral reductions; and
by computing the coefficients of scalar integrals directly, it removes
the need for tensor reductions.  Britto, Cachazo and
Feng~\cite{BCFUnitarity} showed how to combine generalized unitarity
with a twistor-inspired~\cite{Twistor} use of complex momenta to
express all box coefficients as a simple sum of products of tree
amplitudes.  Forde~\cite{Forde} showed how to extend the technique to
triangle and bubble coefficients.  His method uses a complex
parametrization and isolates the coefficients at specific universal
poles in the complex plane.  It is well suited to analytic
calculation.  Upon trading series expansion at infinity for exact
contour integration via discrete Fourier summation~\cite{BlackHatI},
the method can be applied
to numerical calculation as well, where it is intrinsically stable.
Generalized unitarity also meshes well with the subtraction approach
to integral reduction introduced by Ossola, Papadopoulos and Pittau
(OPP)~\cite{OPP}.  As described in ref.~\cite{BlackHatI}, in
\BlackHat{} we use Forde's analytic method, adapted to a numerical
approach.  We evaluate the boxes first, then the triangles, followed
by the bubbles; the rational terms are computed separately.  For each
term computed by cuts, we enhance the numerical stability of Forde's
method by subtracting prior cuts.  This is similar in spirit to,
though different in details from, one aspect of the OPP approach, in
which all prior integral coefficients are subtracted at each stage.

The terms $R_n$, which are purely rational in the spinor variables,
cannot be computed using four-dimensional unitarity methods.  At
present, there are two main choices for computing these contributions
within a process-nonspecific numerical program: on-shell recursion,
and $D$-dimensional unitarity.  Loop level on-shell
recursion~\cite{Bootstrap, Genhel} is based on the tree-level on-shell
recursion of Britto, Cachazo, Feng and Witten~\cite{BCFW}.  The
utility of $D$-dimensional
unitarity~\cite{DdimUnitarity,OneLoopReview,
DdimUnitarityRecent,GKM,OPPrat,ggttg} grows out of the original
observation~\cite{VanNeerven} by van~Neerven that dispersion integrals
in dimensional regularization have no subtraction ambiguities.
Accordingly the unitarity method in $D$ dimensions retains all
rational contributions to amplitudes~\cite{DdimUnitarity}.  This
version of unitarity, in which tree amplitudes are evaluated in $D$
dimensions, has been used in various
analytic~\cite{DdimUnitarityRecent} and
numerical~\cite{GKM,OPPrat,ggttg,CutTools,GZ,W3EGKMZ,OtherLargeN,HPP}
studies.  We have implemented on-shell recursion in \BlackHat{}, along
with a ``massive continuation'' approach --- related to
$D$-dimensional unitarity --- along the lines of Badger's
method~\cite{Badger}. We speed up the on-shell recursion by explicitly
evaluating some spurious poles analytically.  Both approaches are used
for our evaluation of the \Wjjj-jet virtual matrix elements. For
producing the plots in this paper, we use on-shell recursion for the
computation of primitive amplitudes with all negative helicities
adjacent.  These amplitudes have a simple pattern of spurious
poles~\cite{Genhel} (poles which cancel between the cut part $C_n$ and
rational part $R_n$).  For them, on-shell recursion is faster than
massive continuation in the present implementation.

\BlackHat{}'s use of four-dimensional momenta allows it to rely on
powerful four-dimensional spinor techniques~\cite{SpinorHelicity,
  Wplus2ME, TreeReview} to express the solutions for the loop momenta
in generalized unitarity cuts in a numerically stable
form~\cite{BlackHatI}.  In the computation of the rational terms using
on-shell recursion, it also allows convenient choices for the complex
momentum shifts.  In four dimensions one can also employ simple forms
of the tree amplitudes that serve as basic building blocks.  While
spinor methods arise most naturally in amplitudes with massless
momenta, it is straightforward to include uncolored massive external
states such as the $W$ boson~\cite{Wplus2ME}.  The methods are in fact
quite general, and can also be applied usefully to one-loop amplitudes
with internal massive particles, or external massive ones such as top
quarks (treated in the narrow-width
approximation)~\cite{MassiveLoopSpinor,ggttg}.

With the current version of \BlackHat{}, the evaluation of a complete
helicity-summed leading-color virtual interference term for a
two-quark partonic subprocess~(\ref{W3qqggg}),
built out of all the primitive
amplitudes, takes $530$~ms on average for each phase-space point, on a
$2.33$~GHz Xeon processor. The evaluation of a complete four-quark
partonic subprocess~(\ref{W3qqQQg})
with distinct quarks takes $185$~ms (identical quarks take twice as long).
The mix of subprocesses leads to an evaluation
time of 470 ms on average for each phase-space point.  (As described
in \sect{PSISubSection}, in performing the phase-space integration we
sample a single subprocess at each point.)  Using the
``color-expansion sampling'' approach we shall discuss in
\sect{ColorApproximationSection}, evaluating the subleading-color
contributions would multiply this time by about 2.4, giving an average
evaluation time of 1.1~s for the full color calculation.

\subsection{Numerical Stability of Virtual Contributions}
\label{NumStabilitySubsection}

\BlackHat{} computes matrix elements numerically using on-shell methods.
In certain regions of phase space, particularly near the vanishing loci
of Gram determinants associated with the scalar integrals $I_{2,3,4}^i$,
there can be large cancellations between different terms in the
expansion~(\ref{IntegralBasis}) of the cut part $C_n$, or
between the cut part and the rational part $R_n$ in \eqn{CutRational}.
There can also be numerical instabilities in individual terms.
For example, the recursive evaluation of $R_n$ includes a contribution
from residues at spurious poles in the complex plane.  These
residues are computed by sampling points near the pole, in an
approximation to a contour integral which can be spoiled if another
pole is nearby.

In normal operation, \BlackHat{} performs a series of tests to detect
any unacceptable loss of precision.  Whenever \BlackHat{} detects such
a loss, it re-evaluates the problematic contributions to the amplitude
(and only those terms) at that phase-space point using higher-precision
arithmetic (performed by the {\sc QD} package~\cite{QD}).  This
approach avoids the need to analyze in detail the precise origin of
instabilities and to devise workarounds for each case.  It does of
course require that results be sufficiently stable, so that the use of
higher precision is infrequent enough to incur only a modest increase
in the overall evaluation time; this is indeed the case.

The simplest test of stability is checking whether the known infrared
singularity of a given matrix element has been reproduced correctly.
As explained in ref.~\cite{BlackHatI}, this check can be extended
naturally to check individual spurious-pole residues.  Another test
checks the accuracy of the vanishing of certain higher-rank tensor
coefficients.  From the interaction terms in the (renormalizable) QCD
Lagrangian we know on general grounds which high-rank tensor
coefficients have to vanish.  All tensors with rank greater than $m$
must vanish, for the $m$-point integrals with $m=2,3,4$.  If the
integral corresponds to a cut line that is fermionic, then the maximum
rank is reduced by one.  
In our approach the values of the higher-rank tensor coefficients may
be computed without much extra cost in computation time.  For a given
generalized unitarity cut, when using complex loop-momentum parametrizations
along the lines of ref.~\cite{BlackHatI},
these tensor coefficients appear as coefficients of specific monomials
in the complex parameters.  Their values may be extracted as a
byproduct of evaluating the scalar integral coefficients.  Similarly,
in the massive continuation approach to computing the rational terms,
particular tensor coefficients can be associated with specific
monomials in the complex parameters and in an auxiliary complex mass
parameter entering the loop-momentum parametrization.

We apply the latter check when computing coefficients of scalar bubble
integrals, as well as bubble contributions to the rational terms in
the massive continuation approach.  The value of this check is
twofold.  Firstly, it focuses on a small part of the
computation, namely single bubble coefficients. This allows
\BlackHat{} to
recompute at higher precision just the numerically-unstable
contributions, instead of the entire amplitude.  By contrast, the
above-mentioned check of the infrared singularity assesses the
precision of the entire cut part $C_n$ of the given primitive
amplitude, and so it requires more recomputation if it fails.  Secondly,
the check applied to the bubble contributions in the massive
continuation approach assesses the precision of the rational part
$R_n$, which is inaccessible to the infrared-singularity check.

Finally, a further class of tests of numerical precision looks for
large cancellations between different parts of $A_n$,
in particular between $C_n$ and $R_n$ in \eqn{CutRational}.

We have previously assessed the numerical stability of \BlackHat{} for
six-, seven- and eight-gluon one-loop amplitudes~\cite{BlackHatI}, as
well as for the leading-color amplitudes for a vector boson with up to
five partons~\cite{ICHEPBH} used in the present study.  These earlier
studies used a flat phase-space distribution. Here we show the
stability of \BlackHat{} over phase-space points selected in the same
way as in the computation of cross sections and distributions.  As
will be discussed in~\sect{PSISubSection}, the phase-space points are
selected using an integration grid that has been adapted to the
leading-order cross section.

\begin{figure}[tbh]
\includegraphics[clip,scale=0.8]{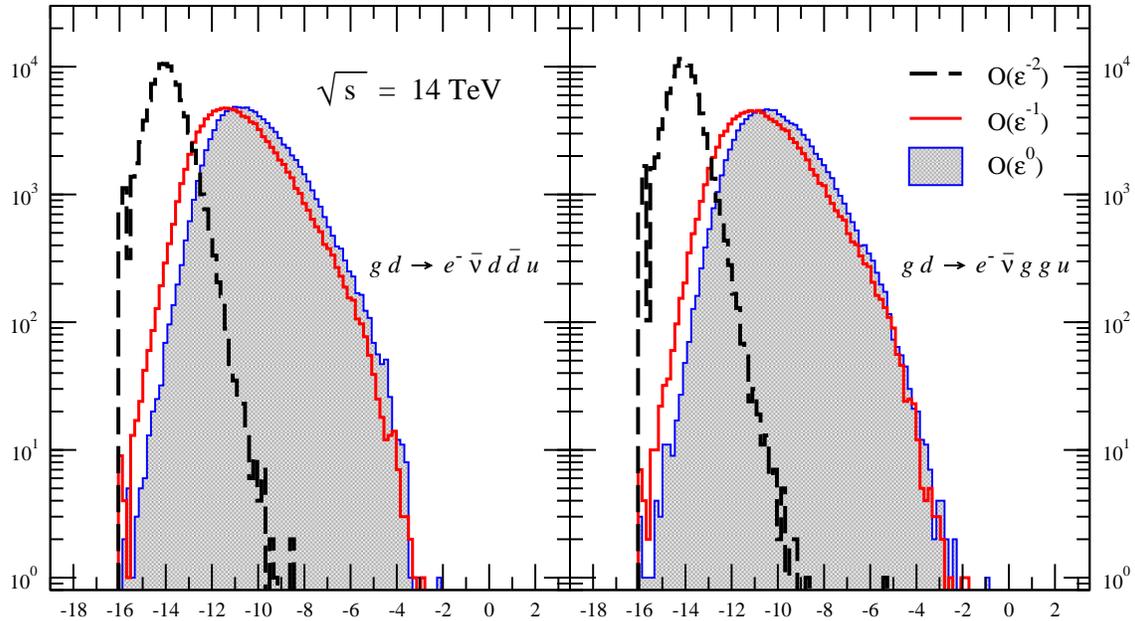}
\caption{The distribution of the relative error in the leading-color
virtual cross section for two subprocesses, $gd \rightarrow e^-\bar\nu
d \bar d u$ and $gd \rightarrow e^-\bar\nu g g u$.  The
phase-space points are selected in the same way as those used to compute
cross sections at the LHC.  The horizontal axis is the logarithm of
the relative error~(\ref{RelError}) between an
evaluation by \BlackHat{}, running in production mode,
and a target expression evaluated using higher precision
with at least $32$ decimal digits (or up to $64$ decimal digits for
unstable points).  The vertical axis shows the number of events out of
100,000 with the corresponding error.  The dashed (black) line shows
the $1/\eps^2$ term; the solid (red) curve, the $1/\eps$ term; and the
shaded (blue) curve, the finite ($\eps^0$) term.}
\label{2q3g2lLHCStabilityFigure}
\end{figure}

In \fig{2q3g2lLHCStabilityFigure}, we illustrate the numerical
stability of the leading-color virtual interference term
(or squared matrix element), 
$d\sigma_V$, summed over colors and over all helicity
configurations for two subprocesses, $gd \rightarrow e^-\bar\nu d \bar
d u$ and $gd \rightarrow e^-\bar\nu g g u$.  (The grid here has been
adapted to each of the subprocesses individually, instead of to the sum
over subprocesses.)  We have checked that the other subprocesses are
similarly stable. The horizontal axis of
\fig{2q3g2lLHCStabilityFigure} shows the logarithmic error,
\begin{equation}
\log_{10}\left(\frac{|d \sigma_V^{\rm num}- d \sigma_V^{\rm target}|}
          {| d \sigma_V^{\rm target}|} \right),
\label{RelError}
\end{equation}
for each of the three components:
$1/\epsilon^2$, $1/\epsilon^1$ and $\epsilon^0$, where $\eps = 
(4-D)/2$ is the dimensional regularization parameter.
The targets have been computed by \BlackHat{}
using multiprecision arithmetic with at least $32$ decimal digits, 
and $64$ if the point is deemed unstable.
The overwhelming majority (99.9\%) of events are computed to better than
one part in $10^4$ --- that is, to the left of the `$-4$' mark on the
horizontal axis.

We have also examined distributions in which each bin is weighted by the
requisite squared matrix element and Jacobian factors.
We find that they have
quite similar shapes to the unweighted distributions shown in
\fig{2q3g2lLHCStabilityFigure}.  This implies that the few events with
a relative error larger than $10^{-4}$ make only a small contribution
to the total cross section.  We have verified that the
difference between normal and high-precision evaluation in the total
cross section, as well as bin by bin for all distributions studied, is
at least three orders of magnitude smaller than the corresponding
numerical integration error.


\subsection{Real-Emission Corrections}

\begin{figure}[tbh]
\includegraphics[clip,scale=0.5]{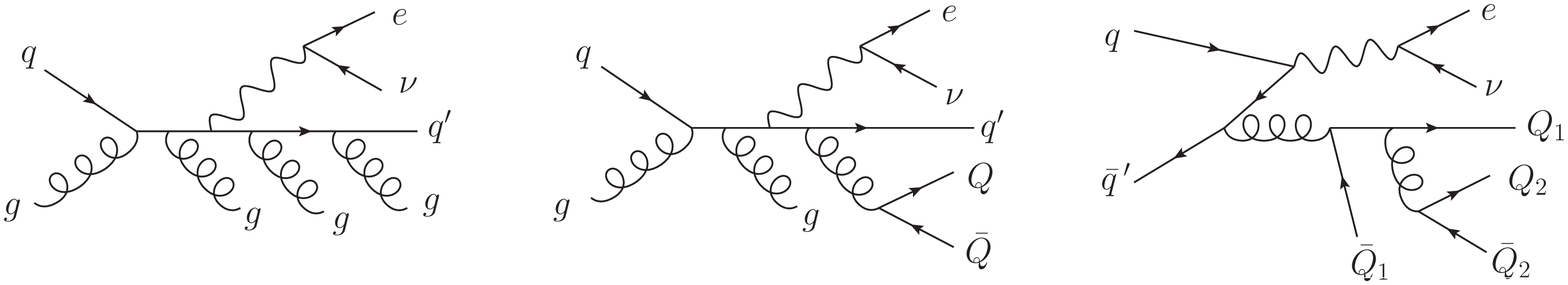}
\caption{Representative diagrams for the eight-point tree-level amplitudes,
$q g \rightarrow e \nu \, q' \! g g g$,
$q g \rightarrow e \nu \, q' \! g Q \bar Q$,
and $q \bar{q}' \rightarrow e \nu \, Q_1 \bar{Q}_1 Q_2 \bar{Q}_2$.
The $e \nu$ pair couples to the quarks via a $W$ boson.}
\label{TreeDiagramsFigure}
\end{figure}

In addition to the virtual corrections to the cross section provided
by \BlackHat, an NLO calculation also requires the real-emission
corrections to the LO process.  These terms arise from tree-level
amplitudes with one additional parton: an additional gluon, or a
quark--antiquark pair replacing a gluon.  Representative real-emission
diagrams are shown in \fig{TreeDiagramsFigure}. Infrared singularities
develop when the extra parton momentum is integrated over phase-space
regions unresolved by the jet algorithm or jet cuts.  The resulting
singular integrals cancel against singular terms in the virtual
corrections, and against counter-terms associated with the evolution
of parton distributions.  As mentioned above, to carry out these
cancellations, we use the Catani--Seymour dipole subtraction
method~\cite{CS} as implemented~\cite{AutomatedAmegic} in the
program~\AMEGIC~\cite{Amegic}, which is part of the \SHERPA{}
framework~\cite{Sherpa}. This implementation of dipole subtraction has
already been tested~\cite{AutomatedAmegic} in explicit comparisons
against the {\sc DISENT} program~\cite{PrivateSeymour}. 

\def\acut{\alpha_{\rm cut}} \def\adipole{\alpha_{\rm dipole}} The
implementation introduces two free parameters, $\acut$ and $\adipole$.
The first, $\acut$, parametrizes the volume of phase space to be cut
out around the soft or collinear singularity.  From an analytic point
of view, $\acut$ could be taken to zero, as the cancellation of
counter-terms against the matrix element's singularities is exact.  In
numerical implementations, however, round-off error can spoil this
cancellation.  Previous studies have shown that the final result is
independent of this cut-off parameter once it is sufficiently
small~\cite{AutomatedAmegic}.  We use $\acut=10^{-8}$.

The second parameter, $\adipole$, characterizes a common modification of
subtraction terms in phase space {\it away\/} from the
singularity \cite{alphadipole}, restricting the
support of a given subtraction term to the vicinity of its
singularity.  This allows the program to compute only a subset
of dipole terms, as many will now be identically zero at a given
phase-space point.
Because the number of dipole terms is large
(scaling as $m^3$ for processes containing $m$ partons),
this reduces the computational burden considerably.  We run our code
with several different values of $\adipole$, and check
the independence of the final result on the value of
$\adipole$ ($0<\adipole\le1$).  For example, the LHC $W^-$ results for
$\adipole = 0.03$ agree with those for $\adipole = 0.01$ to better than
half a percent, within the integration errors.  We have also run a
large number of other lower-statistics checks demonstrating that cross
sections are independent of the choice of $\adipole$.  Our default choice
for the LHC is $\adipole = 0.03$, while for the Tevatron it is
$\adipole = 0.01$.

\subsection{Phase-Space Integration}
\label{PSISubSection}

Along with the automated generation of matrix elements and dipole
terms, \SHERPA{} also provides Monte Carlo integration methods.  The
phase-space generator combines {\it a priori} knowledge about the
behavior of the integrands in phase space with self-adaptive
integration methods.  It employs a multi-channel method in the spirit of
ref.~\cite{MultiChannel}.  Single channels (phase-space
parametrizations) are generated by \AMEGIC{} together with the
tree-level matrix elements. Each parametrization reflects the
structure of a Feynman amplitude, roughly reproducing its resonances,
decay kinematics, and its soft and collinear structure. The most important
phase-space parametrizations, determined by the adapted relative
weight within the multi-channel setup, are further refined using
\Vegas{}~\cite{Vegas}.

The phase-space optimization (adaptation of channel weights and
\Vegas{} grids) is performed in independent runs before the actual
computation starts.  The optimization is done on the sum of all
contributing parton-level processes.  
We refer collectively to all the parameters of the optimization 
as the integration grid.
Separate integration grids are constructed for the LO terms and for the
real-emission contributions.
To integrate the virtual contributions, we re-use the grid constructed
for the LO terms.  This procedure avoids the computational expense of
evaluating the virtual terms merely for grid construction.
The virtual to LO ratio is sufficiently flat across phase space that
this results in only a slight inefficiency when evaluating distributions.

Following the initialization phase, the integration grids are frozen.
In the ensuing production phase, we sample over subprocesses so that
only a single parton-level subprocess is evaluated per phase-space point,
selected with a probability proportional to its contribution to the
total cross section.  We choose to integrate the real-emission terms
over about $10^8$ phase-space points, the leading-color virtual parts
over $2\times 10^6$ phase-space points and the subleading-color
virtual parts over $10^5$ phase-space points.  The LO and
real-subtraction pieces are run separately with $10^7$ points each.
These numbers are chosen to achieve a total integration error of half
a percent or less.  For a given choice of scale $\mu$, they give
comparable running times for the real-emission and virtual
contributions. Running times for leading- and subleading-color virtual
contributions are also comparable.  

\subsection{Couplings and Parton Distributions}
\label{CouplingsSection}

We work to leading order in the electroweak coupling and approximate
the Cabibbo-Kobayashi-Maskawa (CKM) matrix by the unit matrix.  This
approximation causes a rather small change in total cross sections for
the cuts we impose, as estimated by LO evaluations using the full CKM
matrix.  At the Tevatron, the full CKM results are about one percent
smaller than with the unit CKM matrix; the difference is even smaller
at the LHC.  We express the $W$-boson couplings to fermions using the
Standard Model input parameters shown in \tab{ElectroWeakTable}. The
parameter $g_w^2$ is derived from the others via,
\begin{equation}
g_w^2 = {4 \pi \alpha_{\rm QED}(M_Z) \over \sin^2\theta_W}\,.
\end{equation}

\begin{table}
\vskip .4 cm
\begin{tabular}{||c|c||}
\hline
parameter  & value  \\
\hline
$\alpha_{\rm QED}(M_Z)$ & $ 1/128.802 \;$ \\
\hline
$M_W $  & $\; 80.419$ GeV  $\;$  \\
\hline
$\sin^2\theta_W$ & $\; 0.230 \;$ \\
\hline
$\Gamma_W$  & $2.06$ GeV  \\
\hline
$g^2_w$ &  0.4242 (calculated)  \\
\hline
\end{tabular}
\caption{Electroweak parameters used in this work.}
\label{ElectroWeakTable}
\end{table}

We use the CTEQ6M~\cite{CTEQ6M} parton distribution functions (PDFs)
at NLO and the CTEQ6L1 set at LO. The value of
the strong coupling is fixed accordingly,
such that $\alpha_S(M_Z)=0.118$ and $\alpha_S(M_Z)=0.130$
at NLO and LO respectively.  We evolve
$\alpha_S(\mu)$ using the QCD beta function for five massless quark
flavors for $\mu<m_t$, and six flavors for $\mu>m_t$.
(The CTEQ6 PDFs use a five-flavor scheme for all $\mu>m_b$, but we use the
\SHERPA{} default of six-flavor running above top-quark mass;
the effect on the cross section is very small, on the order of one percent
at larger scales.)  At NLO we use two-loop running, and at LO,
one-loop running.

\subsection{Kinematics and Observables}

As our calculation is a parton-level one, we do not apply corrections
due to non-perturbative effects such as those induced by the
underlying event or hadronization.  CDF has studied~\cite{WCDF} these
corrections at the Tevatron, and found they are under ten
percent when the $n^{\rm th}$ jet $E_T$ is below 50~GeV, and under
five percent at higher $E_T$.

For completeness we state the definitions of standard
kinematic variables used to characterize scattering
events. We denote the angular separation of two objects (partons, jets
or leptons) by
\begin{eqnarray}
\Delta R= \sqrt{(\Delta \phi)^2+ (\Delta \eta)^2}\,,
\end{eqnarray}
with $\Delta\phi$ the difference in the
azimuthal angles, and $\Delta\eta$ the difference in the
pseudorapidities. The pseudorapidity $\eta$ is given by
\begin{eqnarray}
\eta= -\ln\left(\tan \frac{\theta}{2}\right)\,,
\end{eqnarray}
where $\theta$ is the polar angle with respect to the beam axis.

The transverse energies of massless outgoing
partons and leptons, $E_T=\sqrt{p_x^2+p_y^2}$,
can be summed to give the total partonic transverse energy,
$\HTpartonic$, of the scattering process,
\begin{equation}
\HTpartonic = \sum_p E_T^p + E_T^e + E_T^{\nu} \,.
\label{PartonicHTdef}
\end{equation}
All partons $p$ and leptons are included in $\HTpartonic$, whether or not
they are inside jets that pass the cuts.  We shall see in later
sections that the variable $\HTpartonic$ represents a good
choice for the renormalization and factorization scale of a given event.
Although the partonic version is not directly measurable,
for practical purposes as a scale choice, it is essentially equivalent
(and identical at LO)
to the more usual jet-based total transverse energy,
\begin{equation}
H_T = \sum_j E_{T,j}^\jet + E_T^e + E_T^{\nu} \,.
\label{PhysicalHTdef}
\end{equation}
The partonic version $\HTpartonic$ has the advantage that it is
independent of the cuts; thus, loosening the cuts will not affect
the value of the matrix element, because a renormalization scale of 
$\HTpartonic$ will be unaffected.
On the other hand, we use the jet-based quantity $H_T$,
which is defined to include only jets passing all cuts,
to compute observable distributions.  Note that for \Wjn-jet
production at LO, exactly $n$ jets contribute to \eqn{PhysicalHTdef};
at NLO either $n$ or $n+1$ jets may contribute.

The jet four-momenta are computed by summing the four-momenta
of all partons that are clustered into them,
\begin{equation}
p^\jet_\mu = \sum_{i\in \jet} p_{i\mu}\,.
\end{equation}
The transverse energy is then defined in the usual way, as the
energy multiplied by the momentum unit vector projected onto the transverse
plane,
\begin{equation}
E_T^\jet = E^\jet \sin\theta^\jet\,.
\end{equation}
The total transverse energy as defined in \eqn{PhysicalHTdef}
is intended to match the experimental quantity, given by the sum,
\begin{equation}
H_T^{\rm exp} = \sum_j E_{T,j}^\jet + E_T^e + \ETslash\,,
\label{ExperimentalHTdef}
\end{equation}
where $\ETslash$ is the missing transverse energy.
Jet invariant masses are defined by
\begin{equation}
M^2_{ij} = (p^\jet_i + p^\jet_j)^2 \,,
\label{JetInvMassdef}
\end{equation}
and the jets are always labeled $i,j=1,2,3,\ldots$ in order
of decreasing transverse energy $E_T$, with $1$ being the
leading (hardest) jet.
The transverse mass of the $W$-boson is computed from the
kinematics of its decay products, $W\rightarrow e\nu_e$,
\begin{eqnarray}
M_T^W=\sqrt{2E_T^e E_T^\nu (1- \cos(\Delta \phi_{e\nu}))}\,.
\end{eqnarray}

\subsection{Checks}

We have carried out numerous checks on our code, ranging from checks
of the basic primitive amplitudes in specific regions of phase space
to overall checks of total and differential distributions against
existing codes.  We have compared our results for the total cross
section for \Wjx-jet production (at a fixed scale $\mu = M_W$)
with the results obtained from running \MCFM{}~\cite{MCFM}.
Because the publicly
available version of \MCFM{} does not allow a cut in $M_T^W$ we
eliminated this cut in the comparison.  (We had previously compared
the matrix elements used in the latter code obtained from
ref.~\cite{Zqqgg}, to the results produced purely numerically in
\BlackHat{}.) Agreement at LO and NLO for \Wj-jet production at the
LHC is good to a per mille level. For \Wjj-jet production, at LO we
find agreement with \MCFM{} within a tenth of a percent, while at NLO,
where the numerical integration is more difficult, we find agreement
to better than half a percent\footnote{This level of agreement holds
only for the most recent \MCFM{} code, version 5.5. We thank John
Campbell and Keith Ellis for assistance with this comparison.  Here we
matched \MCFM{} by including approximate top-quark loop contributions, as
given in ref.~\cite{Zqqgg}, and we adopted \MCFM's electroweak parameter
conventions.}.
We find the same level of agreement at NLO at the Tevatron, using a different
set of cuts\footnote{In performing this comparison, we used a
previous version of \MCFM.  The differences between the two versions
at the Tevatron should be minor.}. 

We have carried out extensive validations of our code at a
finer-grained level.  We have confirmed that the code reproduces the
expected infrared singularities (poles in $\e$) for the primitive
amplitudes and the full color-dressed one-loop
amplitudes~\cite{OneloopIR,CS}.  We have also confirmed that the poles
in $\e$ in the full virtual cross section cancel against those found
in the integrated real-subtraction terms~\cite{AutomatedAmegic}.

We checked various factorization limits, both two-particle (collinear)
and multi-particle poles.  These factorization checks
are natural in the context of on-shell recursion.
This method constructs the rational terms
using a subset of the collinear and multi-particle factorization poles;
the behavior in other channels constitutes an independent cross check.
For the leading-color primitive amplitudes, we verified that
all factorization limits of the amplitudes are correct.
(We also checked that all spurious poles cancel.)
For the subleading-color primitive amplitudes, we verified the
correct behavior as any two parton momenta become collinear.
We also checked at least one collinear limit for each partial amplitude.

We had previously computed the leading-color amplitudes for the
subprocess~(\ref{W3qqggg}) in ref.~\cite{ICHEPBH}.  Ellis {\it et
al.}~\cite{W3EGKMZ} confirmed these values, and also computed the
subleading-color primitive amplitudes.  This evaluation used
$D$-dimensional generalized
unitarity~\cite{DdimUnitarity,DdimUnitarityRecent,GKM}, a
decomposition of the processes in \eqns{W3qqQQg}{W3qqggg} into
primitive amplitudes~\cite{qqggg,Zqqgg}, and the OPP formalism for
obtaining coefficients of basis integrals~\cite{OPP}.  We have
compared the subleading-color primitive amplitudes at a selected
phase-space point to the numerical values reported in
ref.~\cite{W3EGKMZ}, and find agreement, up to convention-dependent
overall phases.  Van Hameren, Papadopoulos, and Pittau (HPP) recently
computed~\cite{HPP} the full helicity- and color-summed virtual cross
section for the subprocess~$u\bar{d} \to W^+ggg$ at another
phase-space point, for an undecayed on-shell $W$ boson and including
(small) virtual top-quark contributions.  They used the OPP formalism
and the {\sc CutTools}~\cite{CutTools} and {\sc HELAC-1L}~\cite{HELAC,HPP}
codes.  We have compared the full squared matrix element to the result
produced by the HPP code, with the top-quark contributions
removed\footnote{We thank Costas Papadopoulos and Roberto Pittau for
providing us with these numbers.}.  We find agreement with their value
of the ratio of this quantity to the LO cross section\footnote{We can
recover an undecayed $W$ by integrating over the lepton phase-space;
that integral in turn can be done to high precision by replacing it with a
discrete sum over carefully-chosen points.}.  We have also found
agreement with matrix-element
results from the same code, allowing the $W$ boson to
decay to leptons, as in our setup.  We give numerical values of the
squared matrix elements for an independent set of subprocesses, evaluated
at a different phase-space point, in an appendix.

As mentioned earlier, we verified that the computed values of the
virtual terms are numerically stable when integrated over grids
similar to those used for computing the cross section and distributions.
We also checked that our integrated results do not depend on
$\adipole$, the unphysical parameter controlling the dipole
subtraction~\cite{alphadipole}, within integration uncertainties.


\extraskip
\section{Tevatron Results}
\label{TevatronResultsSection}

In this section we present next-to-leading order results for \Wjjj-jet
production in $p\pb$ collisions at $\sqrt{s}=1.96$ TeV, the
experimental configuration at the Tevatron.  We decay the $W$ bosons
into electrons or positrons (plus neutrinos)
in order to match the CDF study~\cite{WCDF}.  In
our earlier Letter~\cite{PRLW3BH}, we presented results for the third
jet's transverse energy ($E_T$) distribution as well as the total
transverse energy ($H_T$) distribution.  Those calculations employed a
particular leading-color approximation for the virtual terms~\cite{PRLW3BH}.
As discussed in \sect{ColorApproximationSection},
this approximation is an excellent
one, accurate to within three percent.  In the present paper, we
give complete NLO results for a larger selection of distributions,
including all subleading-color terms.  It would be interesting to compare
the new distributions with experimental results from both CDF and D0, as
they become available.

We use the same jet cuts as in the CDF analysis~\cite{WCDF},
\begin{eqnarray}
  E_T^\jet > 20\, {\rm GeV}\,, \hskip 1 cm \quad   |\eta^\jet| < 2\,.
\label{TeVJetCuts}
\end{eqnarray}
Following ref.~\cite{WCDF}, we quote total cross sections using a
tighter jet cut, $E_T^\jet > 25\,{\rm GeV}$.
We order jets by $E_T$.  Both electron and
positron final states are counted, using the same lepton cuts
as CDF,
\begin{eqnarray}
&&E_T^{e} > 20\, {\rm GeV}\,, \hskip 1 cm  |\eta^e| < 1.1\,, \nn \\
&&\ETsl > 30\, {\rm GeV}\,, \hskip 1 cm  M_T^W > 20\, {\rm GeV}.
\label{TeVLeptonCuts}
\end{eqnarray}
(We replace the $\ETsl$ cut by one on the neutrino $E_T^\nu$.)
CDF also imposes a minimum $\Delta R$
between the charged decay lepton and any jet; the effect of this cut,
however, is undone by a specific acceptance correction~\cite{CDFThesis}.
Accordingly, we do not impose it.

For the LO and NLO results for the Tevatron we use an
event-by-event common renormalization and factorization scale,
set equal to the $W$ boson transverse energy,
\begin{eqnarray}
\mu = E_{T}^{W} \equiv \sqrt{M_W^2 + p_T^2(W)} \,.
\label{ETWscale}
\end{eqnarray}
To estimate the scale dependence we choose five values:
$\mu/2, \mu/\sqrt2, \mu, \sqrt2\mu, 2\mu$.

\begin{table}
\vskip .4 cm
\begin{tabular}{|c||c|c|c|}
\hline
number of jets  & CDF & LO & NLO  \\
\hline
1  & $\; 53.5 \pm 5.6 \;$ & $41.40(0.02)^{+7.59}_{-5.94}$ &
            $\;  57.83(0.12)^{+4.36}_{-4.00} \;$ \\
\hline
2  & $6.8 \pm 1.1$  & $6.159(0.004)^{+2.41}_{-1.58}$ &
            $7.62(0.04)^{+0.62}_{-0.86} $  \\
\hline
3 &  $0.84\pm 0.24$  & $0.796(0.001)^{+0.488}_{-0.276}$
   & $0.882(0.005)^{+0.057}_{-0.138}$  \\
\hline
\end{tabular}
\caption{Total inclusive cross sections, in pb, for \Wjn{} jets produced
at the Tevatron with $W\to e\nu$ and $E_T^{n\rm th\hbox{-}jet} > 25$
GeV, using the experimental cuts of ref.~\cite{WCDF}.  The first
column gives the experimental results as measured by CDF.  The
experimental statistical, systematic and luminosity uncertainties have
been combined in quadrature.  The second column shows LO results, and
the third column the complete NLO results.  In each case, the scale
dependence is quoted in super- and subscripts and the numerical
integration uncertainties in parentheses.}
\label{CDFCrossSectionTable}
\end{table}

The CDF analysis used the {\sc JETCLU} cone algorithm~\cite{JETCLU}
with cone radius $R =0.4$.  This algorithm is not generally infrared
safe at NLO, so we use the seedless cone algorithm
\SISCone~\cite{SISCONE} instead.  Like other cone-type algorithms,
\SISCone{} gives rise to jet-production cross sections that can depend
on an overlap threshold or merging parameter, here called $f$.  No
dependence on $f$ can develop at LO, because such dependence would
require the presence of partons in the overlap of two cones. The
\Wj-jet production cross section likewise cannot depend on $f$ at NLO.
We set this parameter to $0.5$.  (Unless stated otherwise we take this
algorithm and parameter choice as our default.)

 We expect similar results at the
partonic level from any infrared-safe cone algorithm.  For \Wjx-jet
production we have confirmed that distributions using \SISCone{}
are within a few percent of those obtained with \MCFM{} using the
midpoint cone algorithm~\cite{Midpoint}.  (The midpoint algorithm is
infrared-safe at NLO for \Wjx-jet production, but not for \Wjjj-jet
production~\cite{SISCONE}.)  The algorithm dependence of \Wjjj-jet
production at the Tevatron at NLO has also been discussed recently
by Ellis {\it et al.}~\cite{EMZ2}.

In \tab{CDFCrossSectionTable}, we collect the results for the total
cross section, comparing CDF data to the LO and NLO theoretical
predictions computed using \BlackHat{} and \SHERPA{}.  In both cases
these are parton-level cross sections.  Results from more
sophisticated (``enhanced'')
LO analyses incorporating parton showering and matching
schemes~\cite{Matching,MLMSMPR,LOComparison}
may be found in ref.~\cite{WCDF}; however,
large scale dependences still remain. (These calculations make different
choices for the scale variation and are not directly comparable to
the LO parton-level predictions given here.)  As
in the experimental analysis, we sum the $W^-$ and $W^+$ cross
sections, which are identical at the Tevatron (for forward-backward
symmetric acceptance cuts).

\begin{figure*}[tbh]
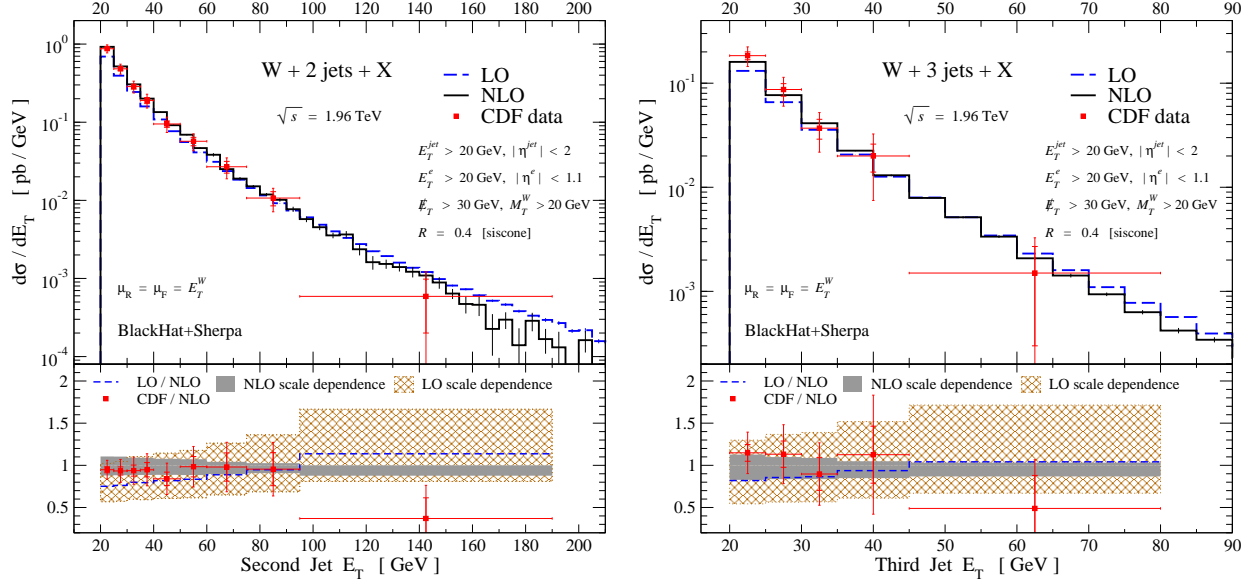

\begin{minipage}[b]{0.49\linewidth}
\includegraphics[clip,scale=0.385]{plots_final/W2jTev_ETWmu_siscone_eb_jets_jet_1_1_Et_2_with_CDF_data.eps}
  \end{minipage}
  \hfill
\begin{minipage}[b]{0.49\linewidth}
\includegraphics[clip,scale=0.385]{plots_final/W3jTev_ETWmu_siscone_eb_jets_jet_1_1_Et_3_with_CDF_data.eps}
\end{minipage}
\caption{The measured cross section, $d \sigma(p\pb \rightarrow e \nu
+ \ge n$-jets)$/dE_T^{n\rm th\hbox{-}jet}$, for inclusive \Wjn-jet
production, compared to full NLO predictions for $n=2,3$.  In the
upper panels the NLO distribution is the solid (black) histogram, and
CDF data points are the (red) points, whose inner and outer error bars,
respectively, denote the statistical and total uncertainties (excluding
the luminosity error) on the measurements added in quadrature.  The LO
predictions are shown as dashed (blue) lines.  The thin vertical lines
in the center of each bin (where visible) give the numerical
integration errors for that bin.  Each lower panel shows the
distribution normalized to the full NLO prediction, using the CDF
experimental bins (that is, averaging over bins in the upper panel).
The scale-dependence bands are shaded (gray) for NLO and cross-hatched
(brown) for LO.}
\label{W23TevatronFigure}
\end{figure*}

We have also computed the \Wjj-jet and \Wjjj-jet total cross sections
at NLO with a larger merging parameter, $f = 0.75$.  (CDF uses a value
of $f=0.75$~\cite{WCDF}, but for a different, infrared-unsafe
algorithm, {\sc JETCLU}.)  The value of the NLO \Wjjj-jet production
cross section of $0.882$~pb in \tab{CDFCrossSectionTable} then
increases to $0.917$~pb (about 4\%).  The \Wjj-jet production cross
section shows a more modest increase from $7.62$~pb to $7.69$~pb
(about 1\%).  Distributions, as for example the ones shown in
\fig{W23TevatronFigure} (see also \tab{CDFE_TCrossSectionTable}),
follow a similar bin-by-bin dependence on $f$.

\begin{table}
\vskip .4 cm
\begin{tabular}{|c||c|c|c|}
\hline
$E_T^{\rm 3rd\hbox{-}jet}$ &
    \multicolumn3{|c|}{$d\sigma/dE_T^{\rm 3rd\hbox{-}jet}$ (pb/GeV)}\\
\hline
  & CDF & LO & NLO  \\
\hline
20-25  & $\;0.184   \pm 0.0394   \;$ & $\; 0.131^{+0.0769}_{-0.0443}\;$
       & $\; 0.160^{+0.0205}_{-0.0277}  \;$
\\\hline
25-30  & $\;0.087   \pm 0.0268   \;$ & $\; 0.066^{+0.0393}_{-0.0224}\;$
       & $\; 0.077^{+0.0075}_{-0.0126}  \;$
\\\hline
30-35  & $\;0.037   \pm 0.0153   \;$ & $\; 0.036^{+0.0216}_{-0.0123}  \;$
       & $\; 0.041^{+0.0036}_{-0.0068}  \;$
\\\hline
35-45  & $\;0.020   \pm 0.0125   \;$ & $\; 0.017^{+0.0103}_{-0.0058}  \;$
       & $\; 0.018^{+0.0009}_{-0.0027}  \;$
\\\hline
45-80  & $\;0.0015  \pm 0.00177   \;$ & $\; 0.0032^{+0.00207}_{-0.00114}\;$
       & $\; 0.0031^{+0.00015}_{-0.00041}  \;$
\\\hline
\end{tabular}
\caption{The differential cross sections, $d \sigma(p\pb \rightarrow e
\nu + \ge 3$-jets)$/dE_{T}^{3\rm rd\hbox{-}jet}$, for \Wjjj{}-jet
production at the Tevatron using the experimental cuts
(\ref{TeVJetCuts}) and (\ref{TeVLeptonCuts}) of ref.~\cite{WCDF}. This
table corresponds to the values plotted in \fig{W23TevatronFigure}.}
\label{CDFE_TCrossSectionTable}
\end{table}

In \fig{W23TevatronFigure}, we compare the $E_T$ distribution of the
second- and third-most energetic jets
in CDF data~\cite{WCDF} to the NLO predictions for
\Wjj-jet and \Wjjj-jet
production, respectively.  For convenience, in
\tab{CDFE_TCrossSectionTable} we collect the data used to construct
the third-jet $E_T$ plot in \fig{W23TevatronFigure}.  We include
scale-dependence bands obtained as described above.\footnote{We
emphasize that the scale-uncertainty bands are only rough estimates of
the theoretical error, which would properly be given by the difference
between an NLO result and one to higher order
(next-to-next-to-leading order).}
The experimental statistical and systematic uncertainties
(excluding an overall luminosity uncertainty of 5.8\%)
have been combined in quadrature.  The upper panels of
\fig{W23TevatronFigure} show the distribution itself, while the lower
panels show the ratio of the LO value and of the data to the NLO
result for the central value of $\mu = \ETW$.
Note that we normalize here to the NLO result, not to LO as 
done elsewhere.  The LO/NLO curve in the bottom panel represents
the inverse of the so-called $K$ factor (NLO to LO ratio).

We do not include PDF uncertainties in our analysis.
For \Wjx-jet production at the Tevatron these uncertainties
have been estimated in ref.~\cite{WCDF}.
For these processes, they are smaller than uncertainties associated
with NLO scale dependence at low jet $E_T$, but larger at high $E_T$.

For reference, we also show the LO distributions and corresponding
scale-dependence bands.  The NLO predictions match the data very well,
and uniformly (without any difference in slope) in all but the highest
$E_T$ experimental bin.  The central values of the LO predictions, in
contrast, have different shapes from the data.  In the upper panels,
we have used 5~GeV bins to plot the predictions, and have superposed
the data points, although CDF used different bins in their analysis.
In the lower panel, which shows the ratio of the LO prediction, and of
the data, to the NLO prediction, we have used the experimental bins,
which are wider at higher $E_T$.  A very similar plot was given
previously~\cite{PRLW3BH}, based on a particular leading-color
approximation.  As we discuss in \sect{ColorApproximationSection},
those results differ only slightly from the complete NLO results
presented here.

\begin{figure}[tbh]
\includegraphics[clip,scale=0.4]{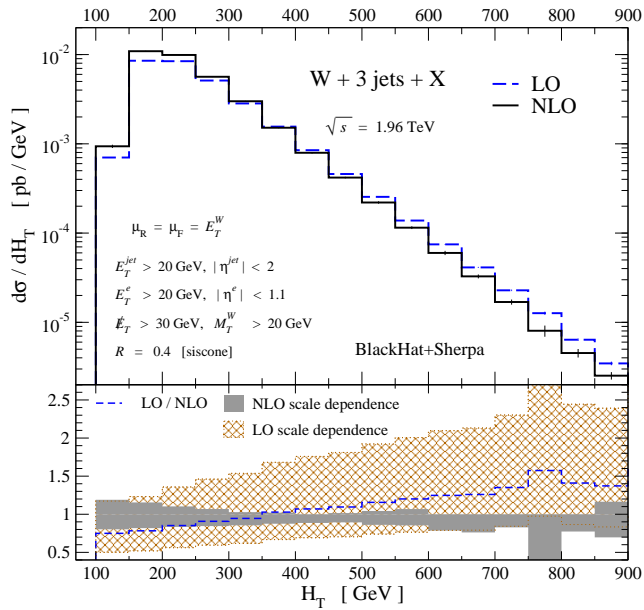}
\caption{Theoretical predictions for the $H_T$ distribution
in \Wjjj-jet production at the Tevatron.
The LO prediction is shown by a dashed (blue) line; the NLO one by a
solid (black) one.  The upper panel shows the distribution itself,
while the lower panel shows the distributions and scale-dependence
bands, cross-hatched (brown) for LO and solid (gray) for NLO,
normalized to the NLO prediction for $\mu = \ETW$.
The numerical integration errors, indicated by thin vertical lines in the
 upper panel, are noticeable only in the tail.}
\label{W3j_HTFigure}
\end{figure}

In \fig{W3j_HTFigure}, we show the distribution for the total
transverse energy $H_T$, given in \eqn{ExperimentalHTdef}.  This
quantity has been used in top-quark studies, and will play an
important role in searches for decays of heavy new particles at the
LHC.  The upper panel shows the LO and NLO predictions for the
distribution, and the lower panel their ratio.  The NLO
scale-dependence band, as estimated using five points, ranges from
$\pm20\%$ around its central value at low $H_T$ 
to $\pm5\%$ around 400~GeV, and back to around
$\pm10\%$ at 800~GeV.  The band is accidentally narrow at
energies near the middle of graph, because the curves associated with
the five $\mu$ values converge as the $H_T$ value rises from lower
values towards the middle ones.  (The fluctuations visible in the tail
of the distribution are a reflection of the limited statistics for the
Monte Carlo integration, as we show a larger dynamical range than in
the $E_T$ spectrum.)  The shape of the LO distribution is noticeably
different, for any of the $\mu$ values, from that at NLO.  At low
$H_T$, the central LO prediction is 20\% below the NLO central value,
whereas at the largest $H_T$ it is nearly 50\% higher.  Thus for
$\mu=E_T^W$ the NLO correction cannot be characterized by a constant
$K$ factor (ratio of NLO to LO results).  We will address some of the
reasons for the difference in shape in the following section.  We note
that the NLO scale-dependence band has a somewhat different appearance
from the corresponding figure in ref.~\cite{PRLW3BH}, because the
latter used wider bins at large $H_T$ and had larger integration
errors.

\begin{figure*}[tbh]
\includegraphics[clip,scale=0.735]{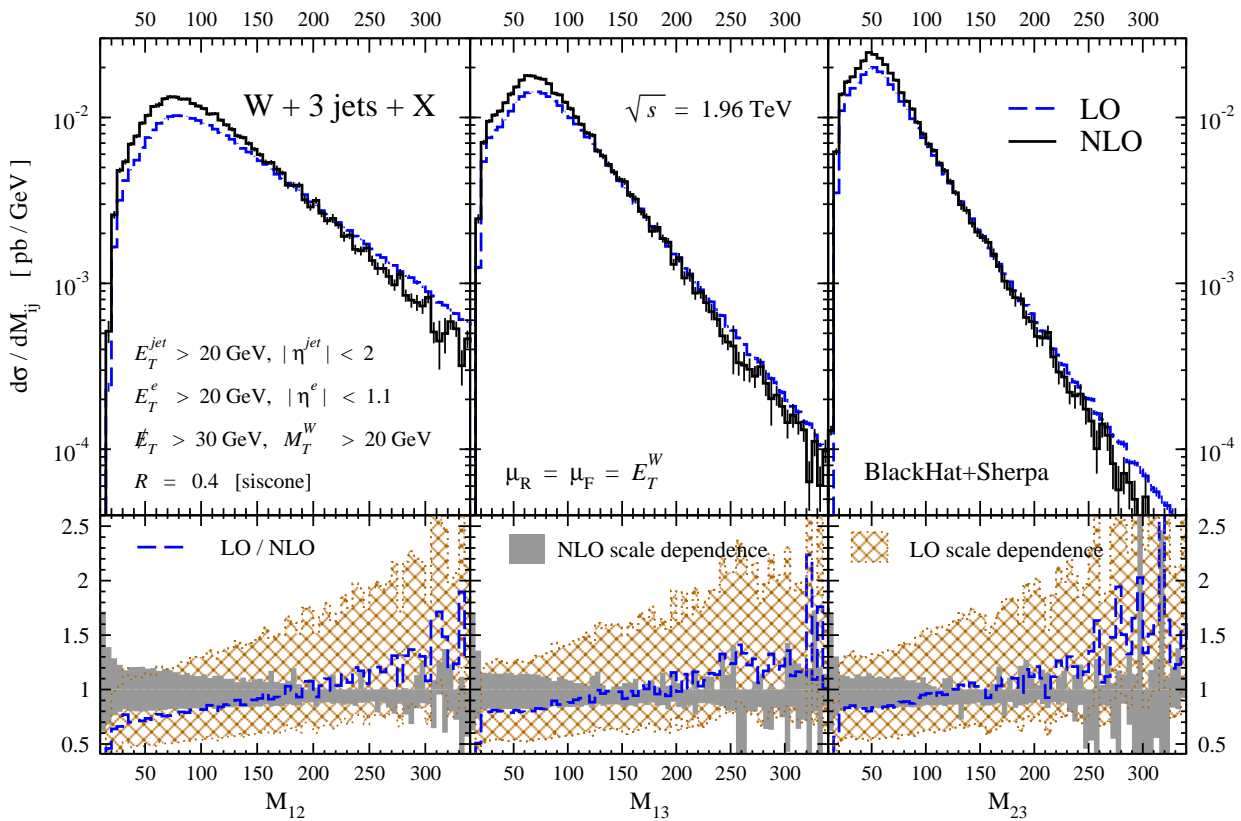}
\caption{LO and NLO predictions for the di-jet invariant mass
distributions $M_{ij}$ (in GeV) in \Wjjj-jet production at the Tevatron.
The histograms and bands have the same meaning as in
\figs{W23TevatronFigure}{W3j_HTFigure}.}
\label{W3j_DJMsFigure}
\end{figure*}

In \fig{W3j_DJMsFigure}, we show the distributions for the three di-jet
invariant masses we can form: hardest and middle jet $M_{12}$,
hardest and softest jet $M_{13}$,
and middle and softest jet $M_{23}$.  The NLO scale-dependence bands are
somewhat broader than for the $E_T$ or $H_T$ distributions.
The distributions become
increasingly steep as we move from masses of hardest to softer jets.
That gross feature is unaltered in passing from LO to NLO,
although each distribution falls off somewhat faster at NLO,
as was the case for $H_T$.


\extraskip
\section{Choosing Scales}
\label{ScaleChoiceSection}

The renormalization and factorization scales are not physical scales.
As such, physical quantities should be independent of them.  They arise in
theoretical calculations as artifacts of defining $\alpha_S$ and the
parton distributions, respectively.  We will follow the usual practice and
choose the two to be equal, $\mu_R = \mu_F = \mu$.
The sensitivity of a perturbative result to the common scale is due to
the truncation of the perturbative expansion; this dependence would be
canceled by terms at higher orders.  NLO calculations greatly reduce
this dependence compared to LO results, but of course do not
eliminate it completely.  In practice, we must therefore choose this
scale.  Intuitively, we would expect a good choice for $\mu$
to be near a ``characteristic'' momentum scale $p$ for the observable
we are computing, in order to minimize logarithms in higher-order terms
of the form $\ln(\mu/p)$.
The problem is that complicated processes such as \Wjjx-jet production
have many intrinsic scales, and it is not clear we can distill them
into a single number.  For any given point in the fully-differential
cross section, there is a range of scales one could plausibly choose.
One could choose a fixed scale $\mu$, the same for all events.
However, because there can be a large dynamic range in momentum
scales (particularly at the LHC, where jet transverse energies well
above $M_W$ are not uncommon), it is natural to pick the scale
$\mu$ dynamically, on an event-by-event basis, as a function of the
observable or unobservable parameters of an event.

A particularly good choice of scale might minimize changes in
shape of distributions from LO to NLO, such as those visible in
\figs{W3j_HTFigure}{W3j_DJMsFigure}.  Such a choice might in turn
make it possible for LO programs incorporating parton showering and
hadronization~\cite{Matching,MLMSMPR,LOComparison} to be more easily
reweighted to reflect NLO results.

\begin{figure*}[tbh]
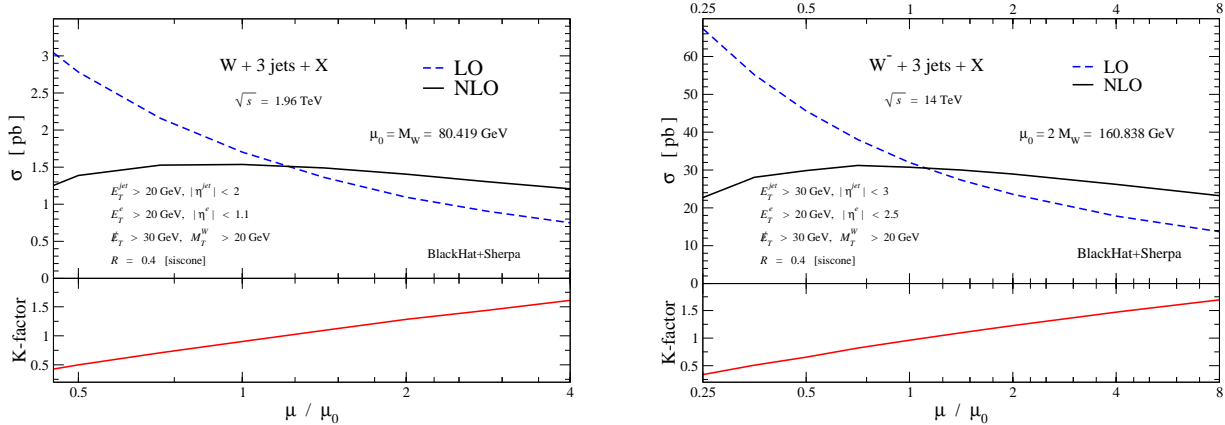

\begin{minipage}[b]{1.\linewidth}
\includegraphics[clip,scale=0.32]{plots_final/W3jTev_FullColor_scale_dependence.eps}
\hskip 1cm
\includegraphics[clip,scale=0.32]{plots_final/Wm3jLHC_FullColor_scale_dependence.eps}
\end{minipage}
\caption{The scale dependence of the cross section for \Wjjj-jet
production at the Tevatron is shown on the left, as a function of the
common renormalization and factorization scale $\mu$, with $\mu_0 =
M_W$.  Similar curves for \Wmjjj-jet production at the LHC are shown
on the right, except that $\mu_0 = 2 M_W$.  In each case, the lower
panel shows the $K$ factor.}
\label{W3ScaleTevLHCFigure}
\end{figure*}

Before turning to dynamical scales and kinematic distributions, let us
first examine how the total cross section depends on a fixed scale.
In \fig{W3ScaleTevLHCFigure} we display this dependence for the
Tevatron\footnote{Note that the Tevatron plot is for $E_T^{3\rm
rd\hbox{-}jet} > 20$ GeV, not the cut $E_T^{3\rm rd\hbox{-}jet} > 25$
GeV used in \tab{CDFCrossSectionTable}.}  and the LHC (left and right
respectively).  We vary the scale between $\mu = M_W/2$ and $4 M_W$
for the Tevatron, and between $M_W/2$ and $16 M_W$ for the LHC.  We
must be careful to vary the scale in a `sensible' range.  For the NLO
calculation in particular, we do not wish to reintroduce large
logarithms of scales.  The figure shows the
characteristic increasing-and-decreasing of the NLO prediction (see
{\it e.g.} refs.~\cite{CTEQscales,CHS}) as well as the monotonicity of
the LO one.  It also shows a substantial reduction in scale dependence
going from LO to NLO.  The lower panels show the $K$ factor.  The
large sensitivity of the LO cross section to the choice of scale
implies a similar large dependence in this ratio.

We thus see that, as expected, the total cross sections at NLO are
much less sensitive to variations of the scale than at LO.  We now
turn to the scale dependence of kinematic distributions.  In this case
the $K$ factor will not only be sensitive to the scale chosen, but it
will in general depend on the kinematic variable.  We will see that a
poor choice of scale can lead to problems not only at LO, but also at
NLO, especially in the tails of distributions.

\begin{figure*}[tbh]
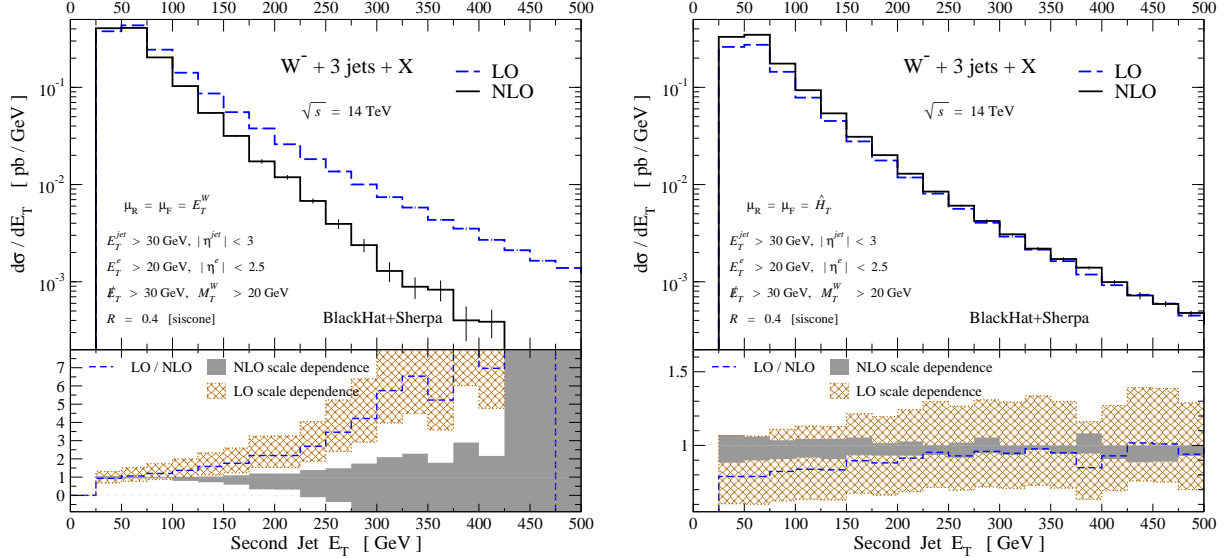

\begin{minipage}[b]{1.\linewidth}
\includegraphics[clip,scale=0.37]{plots_final/Wm3jLHC_ETWmu_siscone_eb_jets_jet_1_1_Et_2.eps}\hskip .5 cm
\includegraphics[clip,scale=0.37]{plots_final/Wm3jLHC_HTmu_siscone_eb_jets_jet_1_1_Et_2.eps}
\end{minipage}
\caption{The $E_T$ distribution of the second jet at LO and NLO,
for two dynamical scale choices, $\mu = E_{T}^W$ (left plot) and
$\mu = \HTpartonic$ (right plot).  The histograms and bands have
the same meaning as in previous figures.
The NLO distribution for $\mu = E_{T}^W$ turns negative
beyond $E_T = 475$ GeV.}
\label{W3jET2ndJetLHCFigure}
\end{figure*}

The sensitivity to a poor scale choice is already noticeable at the
Tevatron, in the shape differences between LO and NLO predictions
visible in \figs{W3j_HTFigure}{W3j_DJMsFigure}. However, it becomes
more pronounced at the LHC because of the larger dynamical range of
available jet transverse energies.  We can diagnose particularly
pathological choices of scale using the positivity of the NLO cross
section: too low a scale at NLO will make the total cross section 
unphysically negative.

This diagnostic can be applied bin by bin in distributions.  For
example, in \fig{W3jET2ndJetLHCFigure} we show the $E_T$ distribution
of the second-most energetic jet of the three, at the LHC.  In the
left plot we choose the scale to be the $W$ transverse energy $E_T^W$
(defined in \eqn{ETWscale}) used earlier in the Tevatron analysis.
Near an $E_T$ of 475~GeV, the NLO prediction for the differential
cross section turns negative!  This is a sign of a poor scale choice,
which has re-introduced large enough logarithms of scale ratios to
overwhelm the LO terms at that jet $E_T$.  Its inadequacy is also
indicated by the large ratio of the LO to NLO distributions at lower
$E_T$, and in the rapid growth of the NLO scale-dependence band with
$E_T$.  In contrast, the right panel of \fig{W3jET2ndJetLHCFigure}
shows that $\HTpartonic$~(defined in \eqn{PartonicHTdef}) provides a
sensible choice of scale: the NLO cross section stays positive, and
the ratio of the LO and NLO distributions, though not completely flat,
is much more stable.

\begin{figure*}[tbh]
\begin{minipage}[b]{0.95\linewidth}
\includegraphics[clip,scale=0.7]{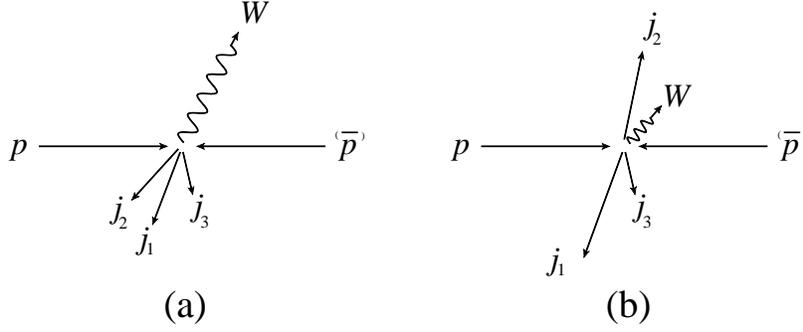}
\end{minipage}
\caption{Two distinct \Wjjj{} jet configurations with rather
different scales for the $W$ transverse energy.
In configuration (a) an energetic $W$ balances the energy
of the jets, while in (b) the $W$ is relatively soft.
Configuration (b) generally dominates over (a) when the jet transverse
energies get large. }
\label{SampleKinFigure}
\end{figure*}

Why is $\mu = E_T^W$ such a poor choice of scale for the
second jet $E_T$ distribution, compared with $\mu=\HTpartonic$?
(For an independent, but related discussion of this question,
see ref.~\cite{Bauer}.)
Consider the two distinct types of \Wjjj{} jet configurations
shown in \fig{SampleKinFigure}.
If configuration (a) dominated, then as the jet $E_T$ increased,
$E_T^W$ would increase along with it, by conservation of transverse
momentum.  However, in configuration (b), the bulk of the transverse
momentum can be balanced between the first and second jet, with
the $W$ and the third jet remaining soft.  In the tail of the
second-jet $E_T$ distribution, configuration (b) is highly favored
kinematically, because it implies a much smaller partonic
center-of-mass energy.  Because $E_T^W$ remains small, the wrong scale
is being chosen in the tail.  Evidence for the dominance of configuration (b)
over (a) in \Wjj-jet production can be found in ref.~\cite{Bauer},
which shows that the two jets become almost back to back as the jet
$E_T$ cut rises past $M_W$.  The negative NLO cross section in
the left panel of~\fig{W3jET2ndJetLHCFigure} provides evidence of the
same domination in \Wjjj-jet production.

However, configuration (b) also tends to dominate in the tails of generic
multi-jet distributions, such as $H_T$ or $M_{ij}$,
in which large jet transverse energies are favored.
The reason is that for jet transverse energies well above $M_W$,
the $W$ behaves like a massless vector boson, and so there is a
kinematic enhancement when it is soft, as in configuration (b).
Exceptions would be in distributions such as the transverse energy
of the $W$ itself, or of its decay lepton, which kinematically
favor configuration (a) in their tails.

\begin{figure*}[tbh]
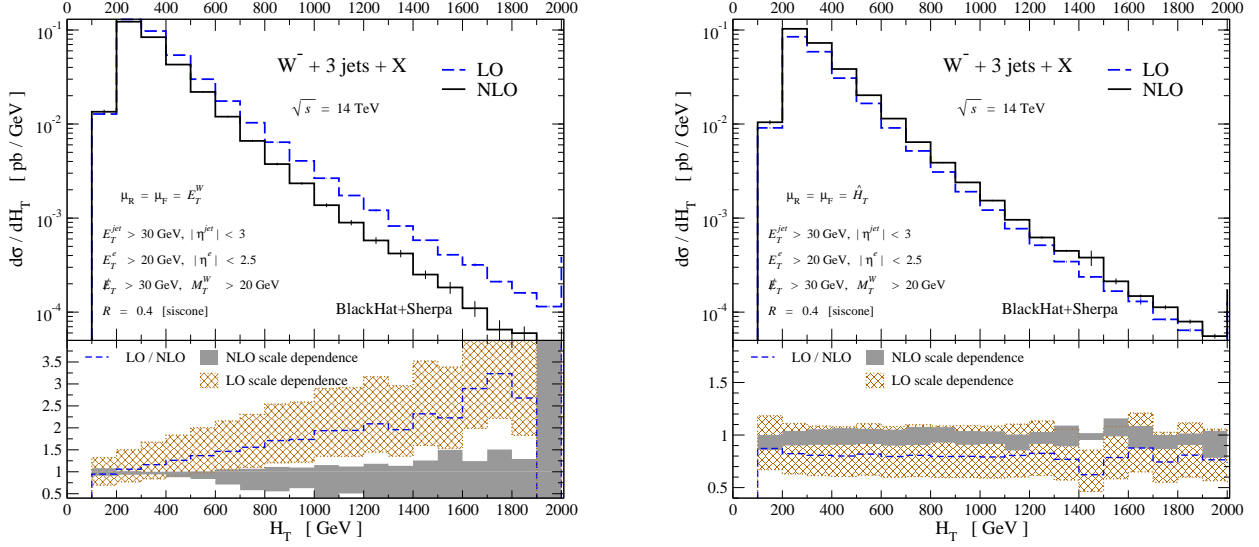

\begin{minipage}[b]{1.\linewidth}
\includegraphics[clip,scale=0.36]{plots_final/Wm3jLHC_ETWmu_siscone_eb_Allp_HT.eps}  \hfill
\includegraphics[clip,scale=0.36]{plots_final/Wm3jLHC_HTmu_siscone_eb_Allp_HT.eps}
\end{minipage}
\caption{The $H_T$ distribution for \Wmjjj-jet production at the LHC.
The scale choices $\mu = E_T^W$ and $\mu = \HTpartonic$ are
shown, respectively, on the left and the right.  The histograms and bands
have the same meaning as in previous figures.}
\label{W3jHTETWmuLHCFigure}
\end{figure*}

\begin{figure*}[bh]
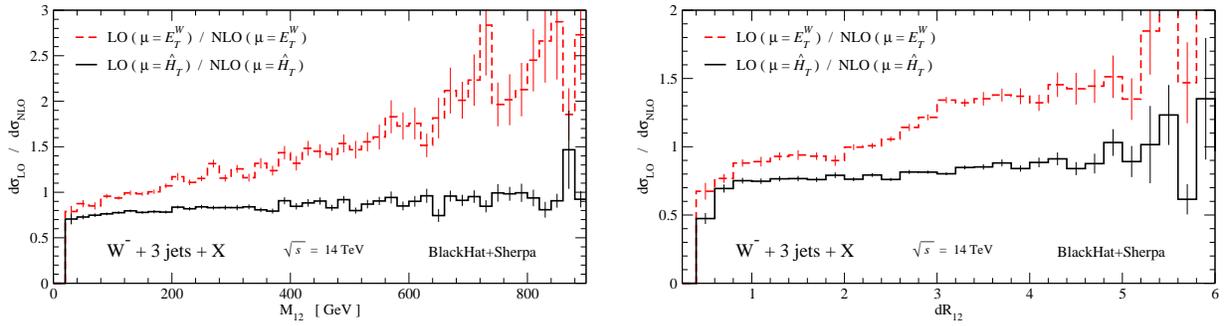

\begin{minipage}[b]{1\linewidth}
\includegraphics[clip,scale=0.33]{plots_final/Wm3jLHC_ETW_HT_lo_over_nlo_central_mu_siscone_jets_jet_1_1_DJM_B1.eps}
\hskip .5 cm
\includegraphics[clip,scale=0.33]{plots_final/Wm3jLHC_ETW_HT_lo_over_nlo_central_mu_siscone_jets_jet_1_1_dR2__A1.eps}
\end{minipage}
\caption{Ratios of LO to NLO predictions for the distributions in the
di-jet invariant mass (left panel) and $\Delta R$ separation (right
panel) for the leading two jets in \Wmjjj-jet production at the LHC.
In each panel, the dashed (red) line gives the scale choice
$\mu=E_T^W$, while the solid (black) line gives the (much flatter)
ratio for $\mu=\HTpartonic$.}
\label{LOvsNLOscale}
\end{figure*}

In contrast to $\mu=E_T^W$, the scale $\mu=\HTpartonic$ becomes large
in the tails of generic multi-jet transverse-energy distributions.
For the distribution of the second jet $E_T$, this is evident from the
close agreement between LO and NLO values, shown in the right panel
of~\fig{W3jET2ndJetLHCFigure}.  The same features are evident, though
less pronounced, in the $H_T$ distributions shown in
\fig{W3jHTETWmuLHCFigure}. The left plot is again for
$\mu=E_T^W$, and the right plot for $\mu=\HTpartonic$.  The shapes of
the LO and NLO distributions for $\mu=E_T^W$ are quite different; the
ratio displayed varies from around 1 at $H_T$ of 200~GeV to around 2
at $H_T$ near 1200~GeV.  In contrast, the ratio for $\mu=\HTpartonic$
is nearly flat.

These features are not special to the $H_T$ distribution itself.  For
example, \fig{LOvsNLOscale} displays the ratio of LO to NLO
predictions for two other \Wjjj-jet distributions for the two scale
choices.  The left panel shows the ratios for the leading di-jet mass,
while the right panel shows ratios for the leading
$\Delta R$ distribution.  Once again the ratios for $\mu=\HTpartonic$
have a much milder dependence than those for $\mu=E_T^W$.

As we shall see further in the next section,
the roughly flat ratio for the choice $\mu=\HTpartonic$
holds for a wide variety of distributions.
It does not hold for all: some NLO corrections cannot be absorbed
into a simple redefinition of the renormalization scale.
The distribution of the second-most energetic jet
in \fig{W3jET2ndJetLHCFigure} provides one example.  A second example,
discussed below,
is the $H_T$ distribution for \Wpjj-jet production in the left plot
of \fig{HTmu_and_HADmuHTFigure}.  A third example (not shown) would
be the $H_T$ distribution for \Wj-jet production; this case is easy
to understand because only configuration (a) (with the second and
third jets erased) is available at LO, while configuration (b) can
dominate at NLO, so effectively a new subprocess opens up at NLO.

Although the $E_T^W$ scale choice is a poor one as far as the tails of
many distributions are concerned, we note that it does give reasonable
results for the Tevatron and LHC total cross sections with our
standard jet cuts, which are dominated by modest jet transverse energies.
For $\mu=E_T^W$, the NLO cross section for \Wmjjj-jet production at the LHC
is $31.37(0.20)^{+0.0}_{-2.47}$ pb, which has much smaller scale variation
than the LO result $37.16(0.07)^{+16.35}_{-10.35}$ pb. (The
parentheses indicate the integration uncertainties,
and subscripts and superscripts the scale variation.)
For $\mu=\HTpartonic$, the NLO value is $27.52(0.14)^{+1.34}_{-2.81}$ pb;
the two NLO results are consistent within the scale variation band.

\begin{figure*}[tbh]
\begin{minipage}[b]{0.95\linewidth}
\includegraphics[clip,scale=0.4]{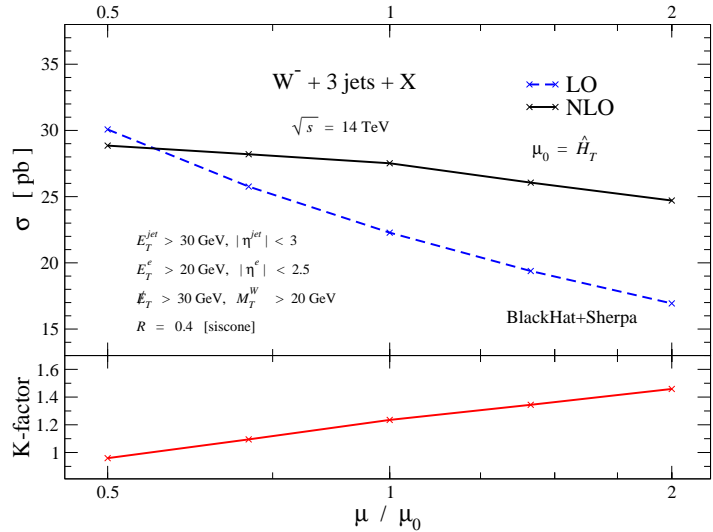}
\end{minipage}
\caption{The scale dependence of the cross section for \Wmjjj-jet
production at the LHC as a function of $\mu/\mu_0$, with
$\mu_0 = \HTpartonic$.
}\label{W3HtScaleLHCFigure}
\end{figure*}

Accordingly, to have a proper description of distributions,
we adopt $\HTpartonic$ as our default choice of scale for
\Wjjj-jet production at the LHC.  In \fig{W3HtScaleLHCFigure} we
display the scale variation of the total cross section, evaluating
it at the five scales $\mu_0/2,
\mu_0/\sqrt2, \mu_0, \sqrt2\mu_0, 2\mu_0$ with $\mu_0 =\HTpartonic$.
As usual, the variation is much smaller at NLO than at LO.
Because $\HTpartonic$ includes a scalar sum, it is somewhat larger than
an ``average'' momentum transfer.
One could choose a scale lower by a fixed ratio,
say $\HTpartonic/2$.  This would shift the
LO-to-NLO ratio curves in \figs{W3jHTETWmuLHCFigure}{LOvsNLOscale},
for example, up towards a ratio of 1.  It would have
only a modest effect on the NLO predictions, however, because the
scale-dependence curve for the NLO cross section is relatively flat.

\begin{figure*}[tbh]
\begin{minipage}[b]{1.\linewidth}
\includegraphics[clip,scale=0.74]{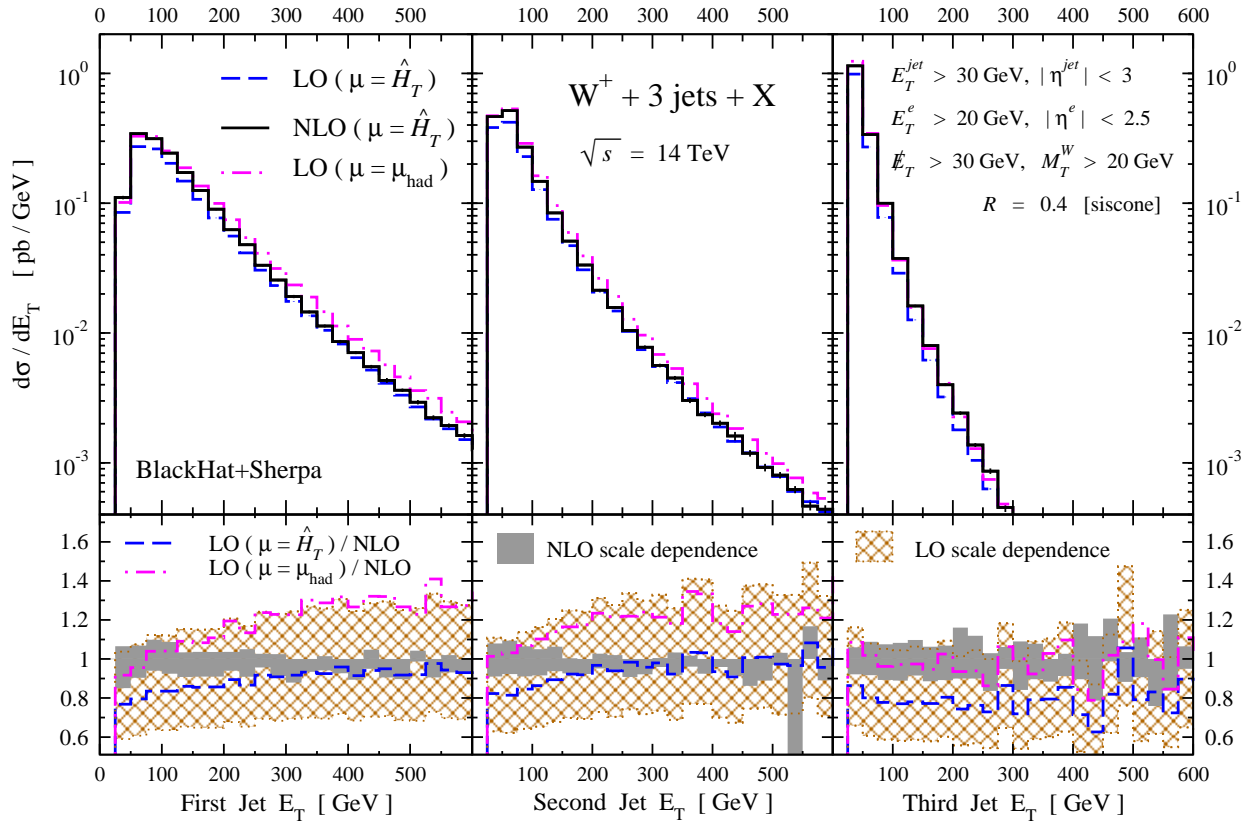}
\end{minipage}
\caption{The first, second, and third jet $E_T$ distributions for
\Wpjjj-jet production.  The scale choice $\mu = \mu_{\rm had}$ at LO,
shown by the dot-dashed (magenta) lines, is compared with the LO and
NLO results using our default scale choice $\mu = \HTpartonic$, shown
respectively in the dashed (blue) and solid (black) lines.  The
scale-dependence bands are shaded (gray) for NLO and cross-hatched
(brown) for LO.  }
\label{ETallHTmu_and_HADmuFigure}
\end{figure*}

It is interesting to compare our default choice $\mu=\HTpartonic$
with the choice of scale advocated
in ref.~\cite{Bauer} on the basis of soft-collinear-effective theory,
\begin{equation}
\mu_{\rm had}^2 = {1\over 4} M_{\rm had}^2 + M_W^2\,.
\label{BauerChoice}
\end{equation}
In this equation $M_{\rm had}$ is the invariant mass of the jets.  (As
explained in ref.~\cite{Bauer}, the factor of $1/4$ is a choice, not
dictated by a principle.)  With this choice, one can greatly reduce
the shift between LO and NLO in \Wjj-jet distributions, compared to
more conventional choices such as $\mu = E_T^W$.  We have confirmed
that for \Wjj-jet production with a few exceptions, such as the decay
lepton transverse energy, the choice $\mu=\mu_{\rm had}$ does fare
better than $\mu=E_T^W$ in bringing LO in line with NLO. How does this
choice fare in \Wjjj{}-jet production?  To answer this question, we
have compared several distributions.  In
\fig{ETallHTmu_and_HADmuFigure}, we consider the $E_T$ distributions
of the first, second and third jets in \Wpjjj-jet production at the
LHC.  We compare the LO results for $\mu = \HTpartonic$ and $\mu =
\mu_{\rm had}$ to the reference NLO results for $\mu =
\HTpartonic$. (Any sensible choice of scale at NLO should give very
similar results.)  As can be seen from the figure, the choice $\mu =
\HTpartonic$ leads to a somewhat flatter LO to NLO ratio than does
$\mu = \mu_{\rm had}$ for the first jet, and performs about as well
for the second and third leading jets. 

\begin{figure*}[tb]
\begin{center}
\begin{minipage}[b]{1\linewidth}
\includegraphics[clip,scale=0.37]{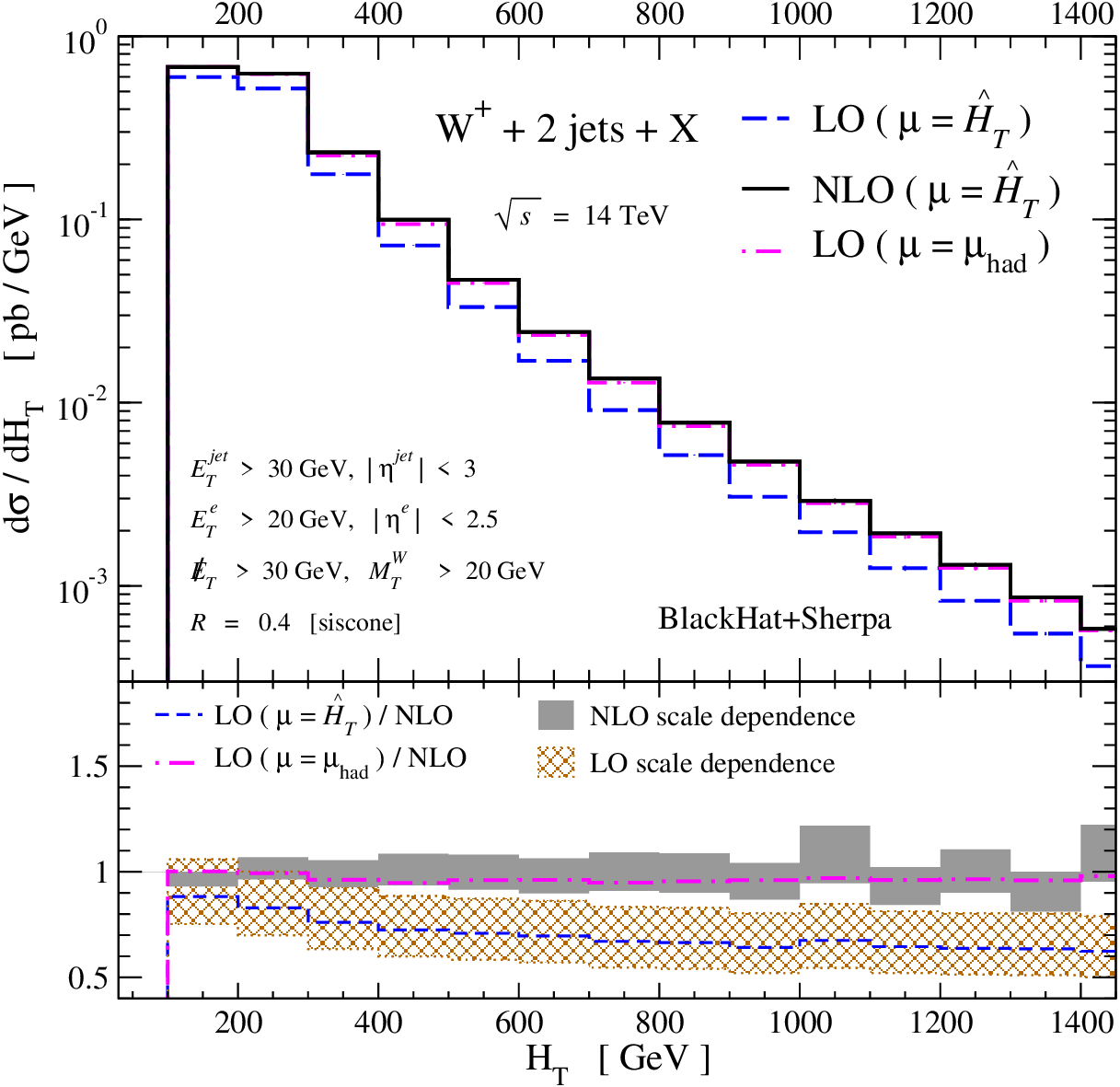}
\hfill
\includegraphics[clip,scale=0.37]{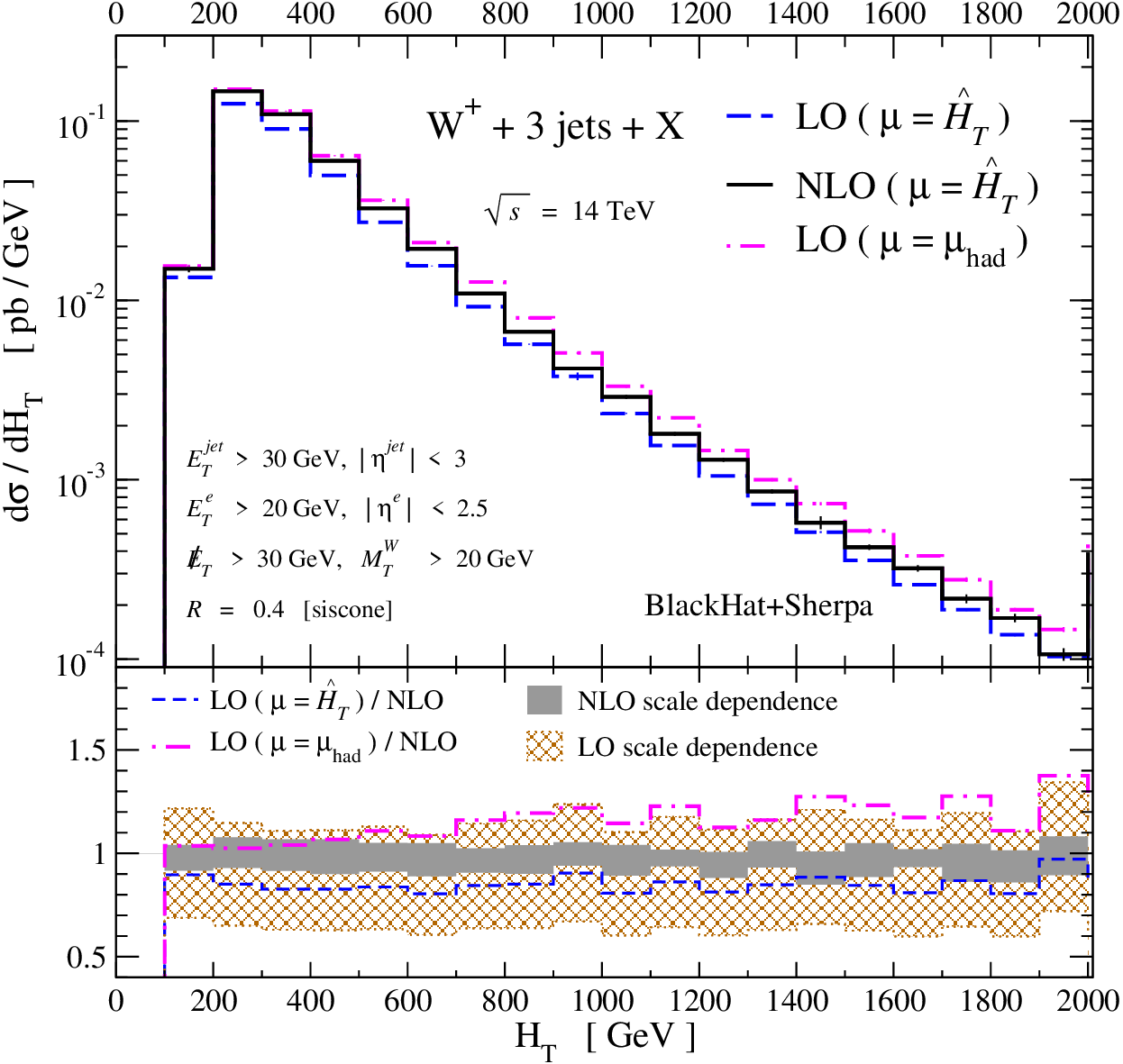}
\end{minipage}
\end{center}
\caption{The $H_T$ distributions for \Wpjjx-jet production.
LO results for $\mu = \mu_{\rm had}$ are compared with
LO and NLO results for $\mu = \HTpartonic$.  The lines and bands
have the same meaning as in \fig{ETallHTmu_and_HADmuFigure}.}
\label{HTmu_and_HADmuHTFigure}
\end{figure*}

It is also instructive to compare the $H_T$ distributions for \Wjj-jet
and \Wjjj-jet production.  The left panel of
\fig{HTmu_and_HADmuHTFigure} shows the distribution in \Wpjj-jet
production; here the scale $\mu=\mu_{\rm had}$ gives an LO result
closer to the NLO one.  On the other hand, in the right panel, which
shows the distribution in \Wpjjj-jet production, the choice
$\mu=\HTpartonic$ gives a LO to NLO ratio which is comparably flat
to the $\mu=\mu_{\rm had}$ choice.

\begin{figure*}[tbh]
\begin{minipage}[b]{1.\linewidth}
  \includegraphics[clip,scale=0.38]{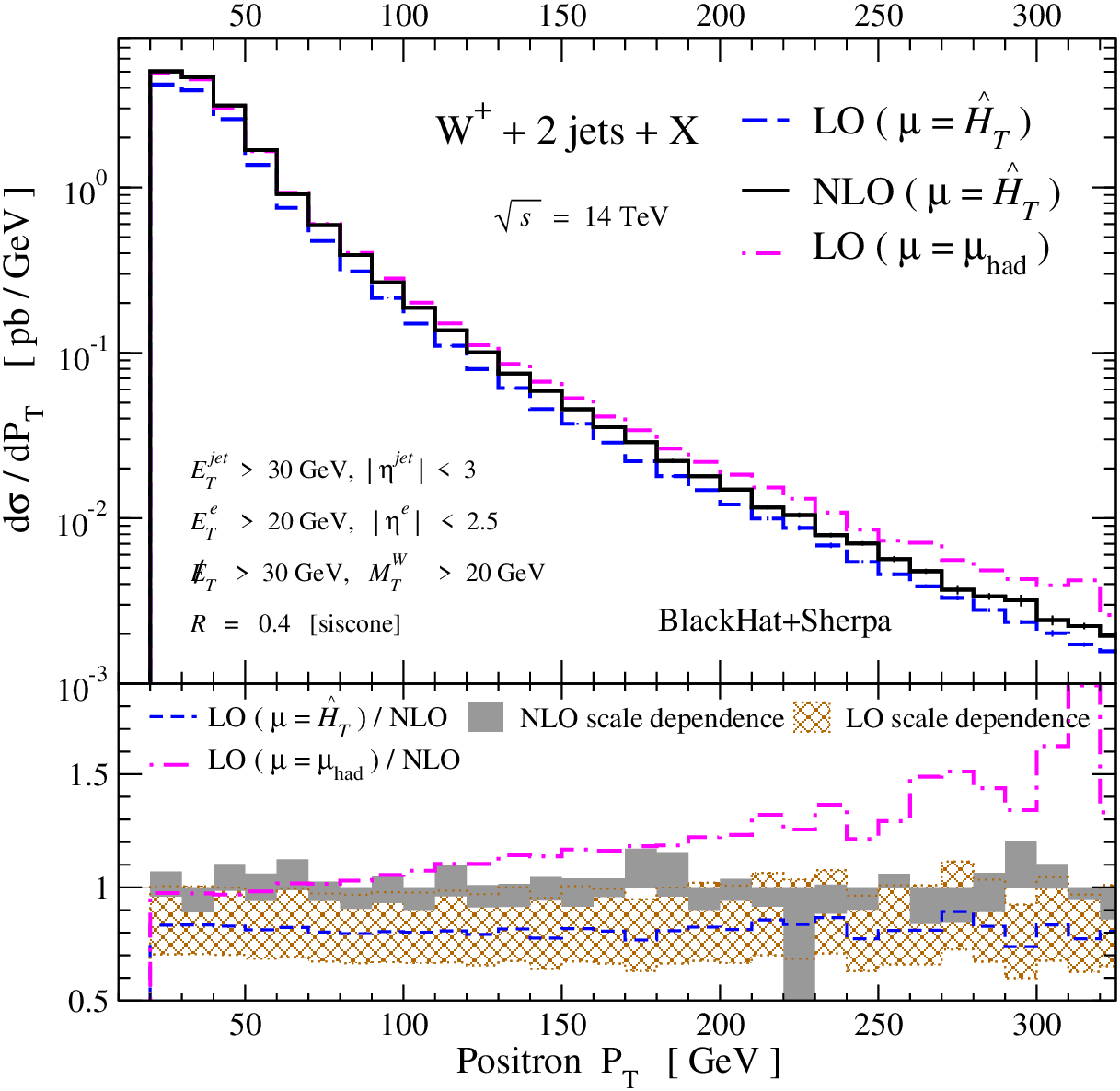}\hfill
  \includegraphics[clip,scale=0.38]{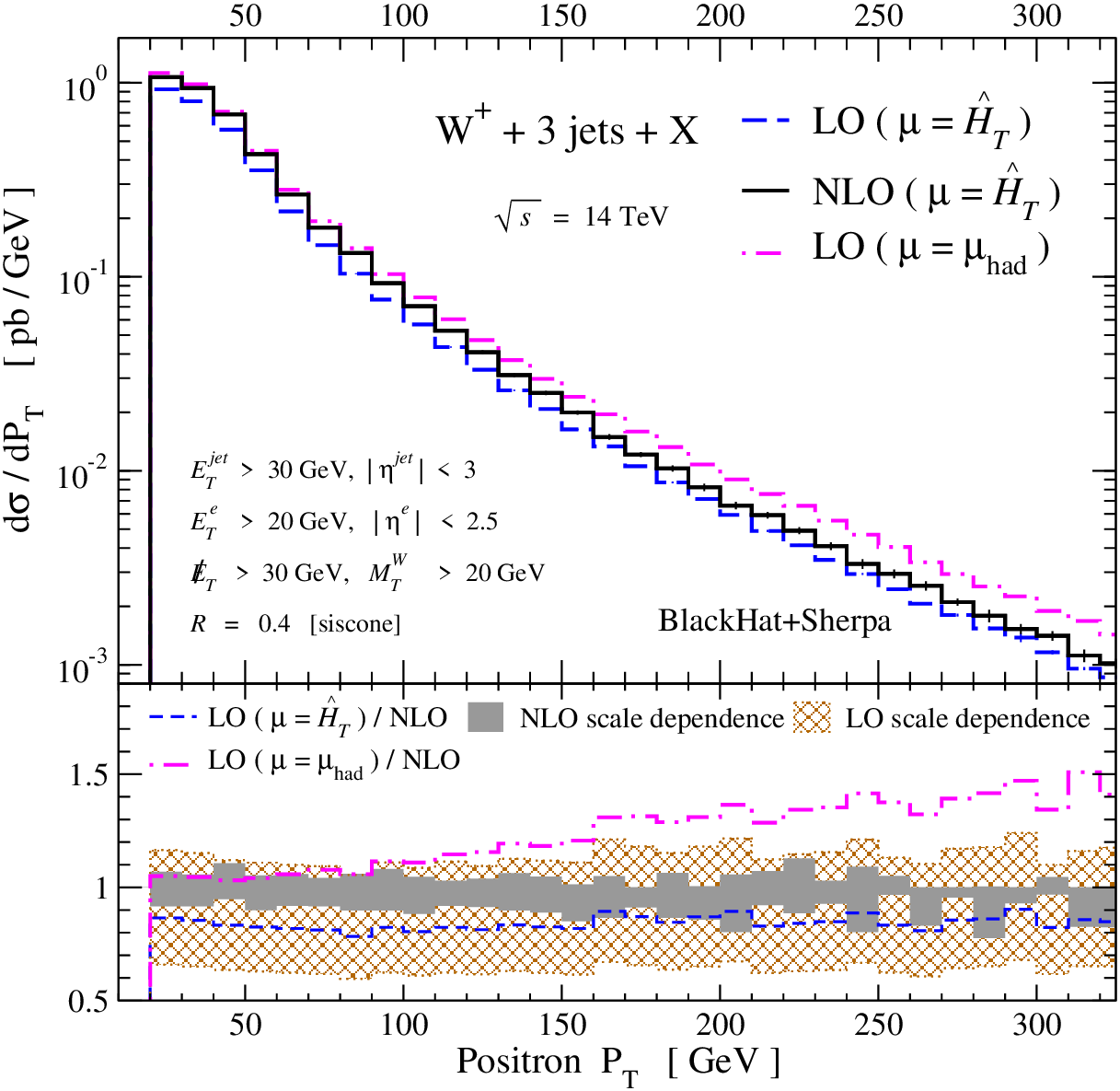}
\end{minipage}
\caption{The distribution in the positron transverse momentum
for \Wpjjx-jet production at the LHC.
LO results for $\mu = \mu_{\rm had}$ are compared with
LO and NLO results for $\mu = \HTpartonic$.  The lines and bands
have the same meaning as in \fig{ETallHTmu_and_HADmuFigure}.}
\label{LeptonDistributions}
\end{figure*}

In contrast, examine the positron $p_T$ (or $E_T$) distribution,
shown in \fig{LeptonDistributions} for \Wpjjx-jet production at the LHC.
As can be seen in the lower panel of each plot,
the choice $\mu = \HTpartonic$ performs better than
$\mu = \mu_{\rm had}$ at LO, in matching the more accurate NLO result
at large values of $E_T^e$.  The reason is that large $E_T^e$ forces
the $W$ transverse energy to be large, which in turn favors configuration (a)
in \fig{SampleKinFigure}, in which a relatively low-mass cluster of jets
recoils against the $W$ boson.  Thus the scale $\mu = \mu_{\rm had}$
drops below the typical momentum transfer in the process.

In summary, both $\mu = \mu_{\rm had}$ and $\mu = \HTpartonic$
are a great improvement over the scale choice $\mu = E_T^W$.
For some distributions $\mu = \mu_{\rm had}$ is a somewhat better
choice at LO than $\mu = \HTpartonic$, while for other 
distributions $\mu = \HTpartonic$ is better. These attributes
should not come as a surprise, given the multi-scale nature 
of jet production.


\extraskip
\section{Predictions for the LHC}
\label{LHCSection}

In this section we present the first complete NLO predictions for
\Wjjj-jet production at the LHC.  The initial run of the LHC will
almost certainly not be at its full design energy of 14~TeV, but we
choose this energy to simplify comparisons to earlier studies.  Most
of the features visible at 14~TeV would of course remain at the lower
energy, such as 10~TeV, of an initial run. The production of
\Wjjj{} jets at the LHC was also studied at NLO in ref.~\cite{EMZ},
however with a set of subprocesses accounting for only 70\% of the
cross section; for on-shell $W$ bosons; and with a less accurate
leading-color approximation than that of ref.~\cite{PRLW3BH}.  For our
analysis of \Wjjj{}-jet production at the LHC, we use the following
kinematical cuts,
\begin{eqnarray}
&& |\eta^\jet| < 3\,, \hskip 1.5 cm
R= 0.4 \,, \hskip 1.5 cm
|\eta^e| < 2.5 \,, \hskip 1.5 cm
E_T^e > 20 \, \hbox{GeV} \,, \hskip 1.5 cm \nn \\
&& \hskip 3 cm
E_T^\nu > 30\, \hbox{GeV} \,, \hskip 1.5 cm
M_T^W > 20\, \hbox{GeV} \,. \hskip 1.5 cm
\label{LHCCuts}
\end{eqnarray}
We also quote total cross sections with both of the 
following jet cuts
\begin{equation}
E_T^\jet > 30\, {\rm GeV} \hskip 7mm \hbox{and}\hskip 7mm
 E_T^\jet > 40 \, {\rm GeV} \,.
\label{LHCJetCut}
\end{equation}
We show distributions only using the first of these two cuts.  We
employ the \SISCone{} jet algorithm~\cite{SISCONE} everywhere (with
$f$ parameter set to $0.5$), except for tables~\ref{WmktTable}
and~\ref{WpktTable} where we use the $k_T$
algorithm~\cite{KTAlgorithm}.

For the LHC we adopt the default
factorization and renormalization scale choices,
\begin{equation}
\mu = \HTpartonic \,,
\end{equation}
where $\HTpartonic$ is defined in \eqn{PartonicHTdef}.
As discussed in the previous section, this choice
does not have the shortcomings of $\mu=E_T^W$ in describing
the large transverse energy tails of generic distributions.

\def\hs{\hskip .05 cm}
\begin{table}[tbp]
\begin{tabular}{|c||c|c||c|c|}
\hline
Number of jets  & \hs LO \hs
            & \hs NLO \hs
            & \hs LO  \hs
            & \hs NLO \hs \\
         & \hs  $E_T^\jet>30$ GeV \hs
            & \hs $E_T^\jet>30$ GeV \hs
            & \hs  $E_T^\jet>40$ GeV \hs
            & \hs $E_T^\jet>40$ GeV \hs \\
\hline
1  & $343.29(0.18)^{+15.65}_{-15.43}$
   & $456.60(1.43)^{+16.61}_{-10.10}$
   & $215.68(0.12)^{+12.19}_{-11.33}$ &
     $298.44(0.77)^{+12.75}_{-8.43}$  \\
\hline
2  & $99.78(0.09)^{+20.81}_{-15.60}$
   & $122.71(0.92)^{+5.88}_{-7.41}$
   &  $58.52(0.063)^{+12.49}_{-9.41}$
   &  $72.96(0.54)^{+3.20}_{-4.54}$
 \\
\hline
3 & $22.28(0.04)^{+7.80}_{-5.34}$
  &  $27.52(0.14)^{+1.34}_{-2.81}$
  & $11.012(0.02)^{+3.87}_{-2.67}$
  &  $13.96(0.07)^{+1.03}_{-1.31}$
 \\
\hline
\end{tabular}
\caption{Cross sections for $W^-$ production using the \SISCone{} jet
algorithm, with jet cuts $E_T^\jet>30$~GeV or $E_T^\jet>40$~GeV.  The
remaining cuts are as in \eqn{LHCCuts}.  }
\label{WmSisconeTable}
\end{table}

\def\hs{\hskip .05 cm}
\begin{table}[htbp]
\begin{tabular}{|c||c|c||c|c|}
\hline
Number of jets  & \hs LO \hs
            & \hs NLO \hs
            & \hs LO  \hs
            & \hs NLO \hs \\
         & \hs  $E_T^\jet>30$ GeV \hs
            & \hs $E_T^\jet>30$ GeV \hs
            & \hs  $E_T^\jet>40$ GeV \hs
            & \hs $E_T^\jet>40$ GeV \hs \\
\hline
1  & $469.37(0.32)^{+21.86}_{-21.26}$
   & $615.77(2.04)^{+23.76}_{-14.39}$
   & $301.20(0.22)^{+17.06}_{-15.86}$
   & $415.50(1.90)^{+19.40}_{-12.86}$ \\
\hline
2  & $143.91(0.18)^{+29.92}_{-22.43}$
   & $174.28(0.48)^{+6.56}_{-10.37}$
   & $86.32(0.12)^{+18.33}_{-13.81}$
   & $105.99(0.31)^{+5.36}_{-5.82}$
 \\
\hline
3 & $34.75(0.05)^{+12.06}_{-8.31}$
  & $41.47(0.27)^{+2.81}_{-3.50}$
  & $17.64(0.02)^{+6.14}_{-4.25}$
  & $21.76(0.15)^{+1.68}_{-1.86}$
 \\
\hline
\end{tabular}
\caption{Cross sections for $W^+$ production  using \SISCone,
with jet cut $E_T^\jet>30$~GeV
or $E_T^\jet>40$~GeV.
The remaining cuts are as in \eqn{LHCCuts}.
\label{WpSisconeTable} }
\end{table}

\def\hs{\hskip .05 cm}
\begin{table}[htbp]
\begin{tabular}{|c||c|c||c|c|}
\hline
Number of jets  & \hs LO \hs
            & \hs NLO \hs
            & \hs LO  \hs
            & \hs NLO \hs \\
         & \hs  $E_T^\jet>30$ GeV \hs
            & \hs $E_T^\jet>30$ GeV \hs
            & \hs  $E_T^\jet>40$ GeV \hs
            & \hs $E_T^\jet>40$ GeV \hs \\
\hline
1   & $343.29(0.18)^{+15.65}_{-15.43}$
    & $444.75(1.44)^{+15.12}_{-8.85}$
    & $215.68(0.12)^{+12.19}_{-11.33}$
    & $290.44(0.77)^{+11.65}_{-7.55}$
 \\
\hline
2  & $102.88(0.09)^{+21.40}_{-16.05}$
   &  $120.07(0.86)^{+4.19}_{-6.33}$
   &  $59.99(0.06)^{+12.78}_{-9.63}$
   &  $70.85(0.42)^{+2.12}_{-3.87}$
 \\
\hline
3 & $25.84(0.05)^{+8.99}_{-6.17}$
  & $29.29(0.16)^{+0.65}_{-2.32}$
  & $12.78(0.02)^{+4.46}_{-3.09}$
  &  $14.89(0.08)^{+0.59}_{-1.18}$
 \\
\hline
\end{tabular}
\caption{Cross sections for $W^-$ production using the $k_T$ algorithm
($R=0.4$), with jet cut $E_T^\jet>30$~GeV or $E_T^\jet>40$~GeV.  The
remaining cuts are as in \eqn{LHCCuts}.  }
\label{WmktTable}
\end{table}

\def\hs{\hskip .05 cm}
\begin{table}[htbp]
\begin{tabular}{|c||c|c||c|c|}
\hline
Number of jets  & \hs LO \hs
            & \hs NLO \hs
            & \hs LO  \hs
            & \hs NLO \hs \\
         & \hs  $E_T^\jet>30$ GeV \hs
            & \hs $E_T^\jet>30$ GeV \hs
            & \hs  $E_T^\jet>40$ GeV \hs
            & \hs $E_T^\jet>40$ GeV \hs \\
\hline
1   & $469.37(0.32)^{+21.86}_{-21.26}$
    & $600.66(2.06)^{+21.83}_{-12.82}$
    & $301.20(0.22)^{+17.06}_{-15.83}$
    & $405.27(1.91)^{+17.91}_{-11.82}$
 \\
\hline
2  &  $148.46(0.19)^{+30.78}_{-23.08}$
   &  $171.45(0.50)^{+3.81}_{-9.39}$
   &  $88.48(0.12)^{+18.75}_{-14.14}$
   &  $103.77(0.31)^{+3.46}_{-5.31}$
 \\
\hline
3 &  $40.27(0.05)^{+13.89}_{-9.59}$
  &  $44.55(0.28)^{+1.59}_{-3.08}$
  &  $20.45(0.03)^{+7.09}_{-4.93}$
  &  $23.20(0.16)^{+0.94}_{-1.67}$
 \\
\hline
\end{tabular}
\caption{Cross sections for $W^+$ production  using the $k_T$ algorithm,
with jet cut $E_T^\jet>30$~GeV
or $E_T^\jet>40$~GeV.
The remaining cuts are as in \eqn{LHCCuts}.
\label{WpktTable} }
\end{table}

At the LHC, a $pp$ collider,
the total rates and the shapes of some distributions are
quite different for $W^-$ and $W^+$ production.
At 14~TeV, the $qg$ initial state accounts for over
half of \Wjn-jet production. There are
considerably more $u$ quarks than $d$ quarks in the proton
in the relevant range of the momentum fraction $x$,
leading to greater production of $W^+$ than $W^-$.
Accordingly, we quote separate results for total cross sections in
tables~\ref{WmSisconeTable}--\ref{WpktTable}.  In
table~\ref{WmSisconeTable}, we show the \Wmjjja-jet cross sections
using the \SISCone{} algorithm, for two different choices of jet $E_T$
cut, 30~and 40~GeV.  The corresponding results for \Wpjjja-jet
production are given in table~\ref{WpSisconeTable}.  In
tables~\ref{WmktTable} and~\ref{WpktTable}, we show the corresponding
results for the $k_T$ jet algorithm with a pseudo-cone radius of 0.4,
for $W^-$ and $W^+$ production respectively.
It is interesting to note that while the NLO cross sections
for \Wjx-jet production are larger for the \SISCone{} algorithm
than for $k_T$ (with the algorithm parameters we have chosen),
the relative size is reversed for \Wjjj-jet production.
(The entries for the LO $W+1$-jet cross section are identical for 
the \SISCone{} and $k_T$ algorithms because the same set of 
events was used to compute them.)

We next describe NLO results for kinematic distributions.
For distributions that do not differ appreciably for $W^-$ and $W^+$
production, except for overall normalization, we generally
show a single distribution.

\begin{figure*}[tbh]
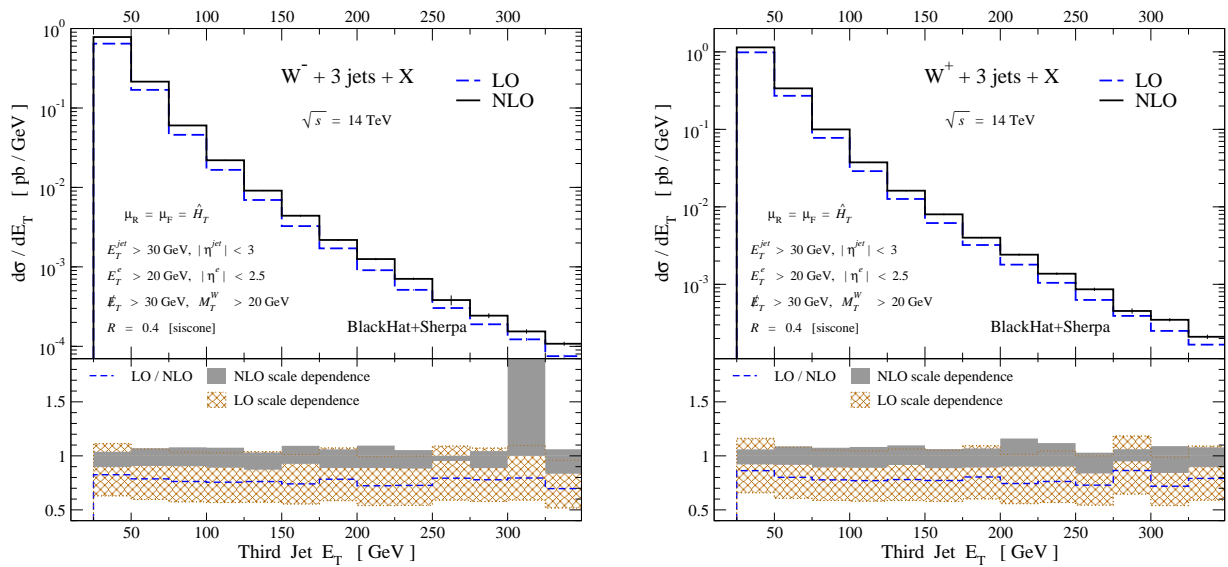

\begin{minipage}[b]{1.\linewidth}
  \includegraphics[clip,scale=0.37]{plots_final/Wm3jLHC_HTmu_siscone_eb_jets_jet_1_1_Et_3.eps} \hskip .8 cm
  \includegraphics[clip,scale=0.37]{plots_final/Wp3jLHC_HTmu_siscone_eb_jets_jet_1_1_Et_3.eps}
\end{minipage}
\caption{ The $E_T$ distributions of the third jet, $d
\sigma(W\rightarrow e \nu + \ge 3$-jets)$/dE_T^{3\rm rd\hbox{-}jet}$,
at the LHC.  The left panel shows the case of $W^-$ and the right of
$W^+$.
}
\label{W3jLHCHTmu_EtFigure}
\end{figure*}

For the inclusive production of \Wjjj{} jets, a basic
quantity to examine is the $E_T$ distribution for the thirdmost
leading jet in $E_T$.  This distribution is shown
in~\fig{W3jLHCHTmu_EtFigure}.  As in the Tevatron results, the scale
uncertainty is considerably reduced at NLO compared to LO.  With our
default choice of scale $\mu = \HTpartonic$, the ratio of LO to NLO
predictions displayed in the lower panels is rather flat over the
entire displayed region.  (The upward spike in the NLO band in
the $W^-$ plot at 300 GeV is due to a statistical fluctuation
in the evaluation at $\mu=\HTpartonic/2$.)
This plot may be compared to the
$E_T$ distribution of the second-most energetic jet
shown in the right panel
of \fig{W3jET2ndJetLHCFigure}, which undergoes significant shape
change between LO and NLO predictions, though less than for the
scale choice $\mu = E_{T}^{W}$.  The dynamic range we show here is larger
than in the corresponding plot for the Tevatron.

\begin{figure*}[tbh]
\begin{minipage}[b]{1.\linewidth}
  \includegraphics[clip,scale=.4]{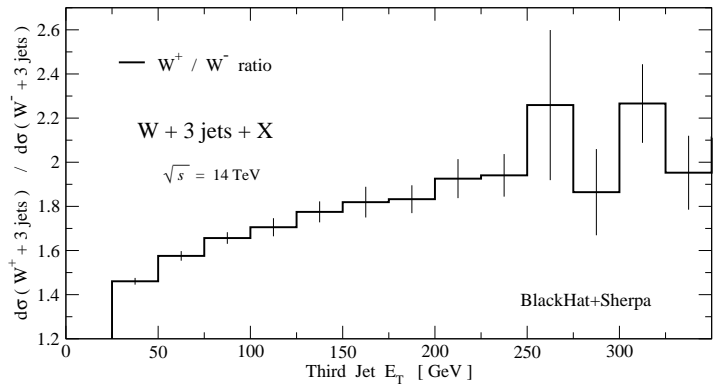}
\end{minipage}
\caption{The ratio of the $E_T$ distribution of the third jet accompanying
a $W^+$ to the same distribution for $W^-$, evaluated at NLO using
$\mu=\HTpartonic$.  The two distributions are shown separately in
\fig{W3jLHCHTmu_EtFigure}.  The thin vertical lines denote the 
numerical integration errors.}
\label{W3jLHCHT_Et_ratioFigure}
\end{figure*}

In order to examine shape differences between the $E_T$ distributions
in $W^+$ and $W^-$ production, in \fig{W3jLHCHT_Et_ratioFigure}
we show the ratio of the two distributions plotted in
\fig{W3jLHCHTmu_EtFigure}.  The ratio
is greater than unity at low $E_T$ due to the larger total cross
section for $W^+$ production compared to $W^-$, as given in
tables~\ref{WmSisconeTable} and~\ref{WpSisconeTable}.  The ratio
increases significantly with $E_T$, on the order of 25 percent over
the range of the plot, because larger $E_T$ forces larger partonic
center-of-mass energies, and hence larger values of $x$ where the $u$
quark distribution is more dominant. 

The $H_T$ distribution also has slightly different shapes for
$W^-$ and $W^+$ production.  The right panel of
\fig{W3jHTETWmuLHCFigure} shows the $H_T$ distribution in 
$W^-$ production (with $\mu = \HTpartonic$).
The corresponding plot for $W^+$ is given in the right panel of
\fig{HTmu_and_HADmuHTFigure}.  Across the displayed range, the ratio of
the  NLO $W^+$ to $W^-$ distributions (not shown)
increases slightly.
The increase occurs for the same reason as the third jet $E_T$ distribution.

\begin{figure*}[tbh]
\begin{minipage}[b]{1.\linewidth}
\includegraphics[clip,scale=0.74]{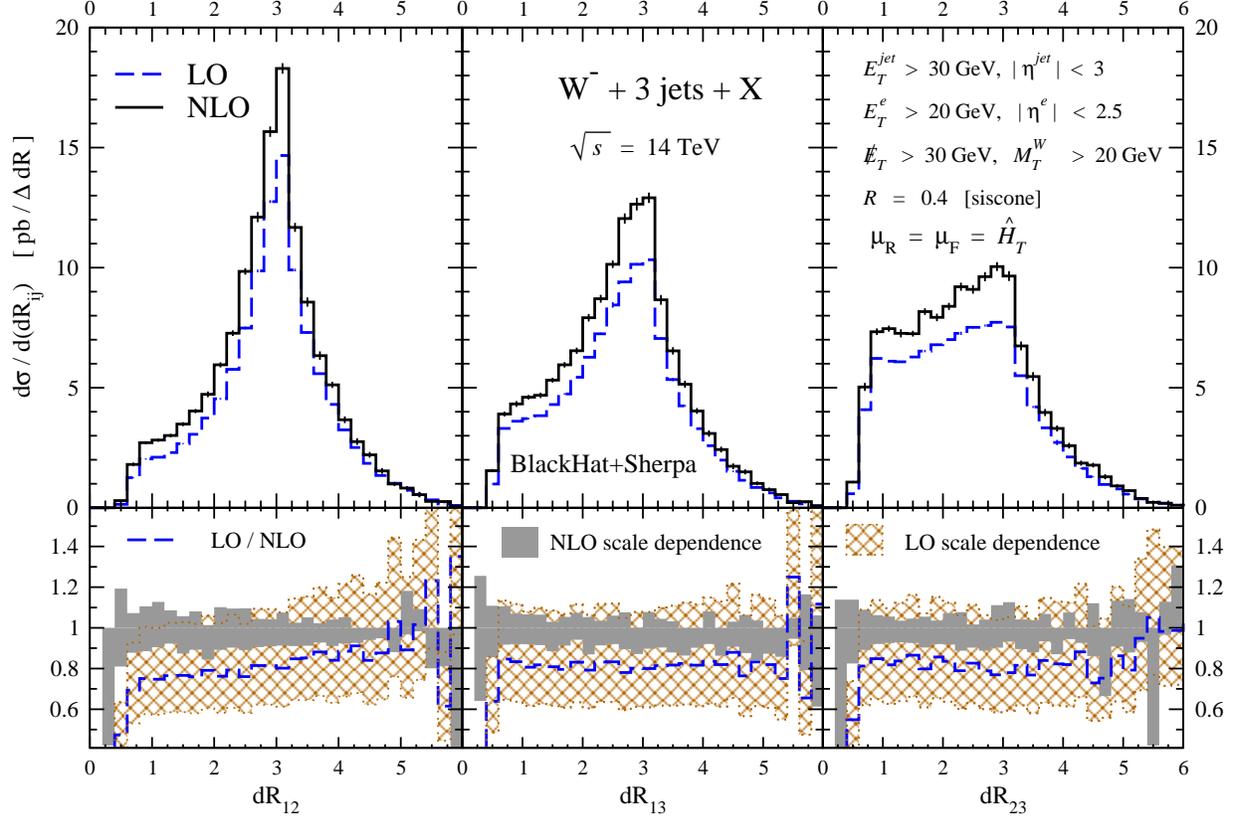}
\end{minipage}
\caption{The $\Delta R$ distributions for \Wmjjj-jet production, from
left to right: between the first and second, first and third, and
second and third jets, using the scale choice $\mu = \HTpartonic$.
The $\Delta R_{12}$ distribution between the first and second jets
shows a significant shape change in going from LO to NLO, while the
other two cases are flat.  The distributions for \Wpjjj-jet production
are similar.}
\label{Wm3jLHCHT_dRFigure}
\end{figure*}

\Fig{Wm3jLHCHT_dRFigure} shows the differential distributions with
respect to di-jet separations $\Delta R_{ij}$.  The two hardest jets,
labeled 1 and 2, are more likely to be produced in a back-to-back
fashion, leading to a more peaked distribution around $\pi$.  As in
other distributions, the NLO scale-dependence band is much smaller
than the LO one.  The LO and NLO distributions for the separation of
the leading two jets are somewhat different from each other in shape.
This is presumably due to the effect of additional radiation allowing
kinematic configurations where the jets are closer together, thereby
pushing the weight of the distribution to smaller $\Delta R$ values,
although the position of the peak is essentially unchanged.
The shapes of the other
two distributions are similar at LO and NLO.  All three distributions
show sizable shifts in their overall normalization, for
$\mu=\HTpartonic$.

\begin{figure*}[tbh]
\begin{minipage}[b]{1.\linewidth}
  \includegraphics[clip,scale=0.735]{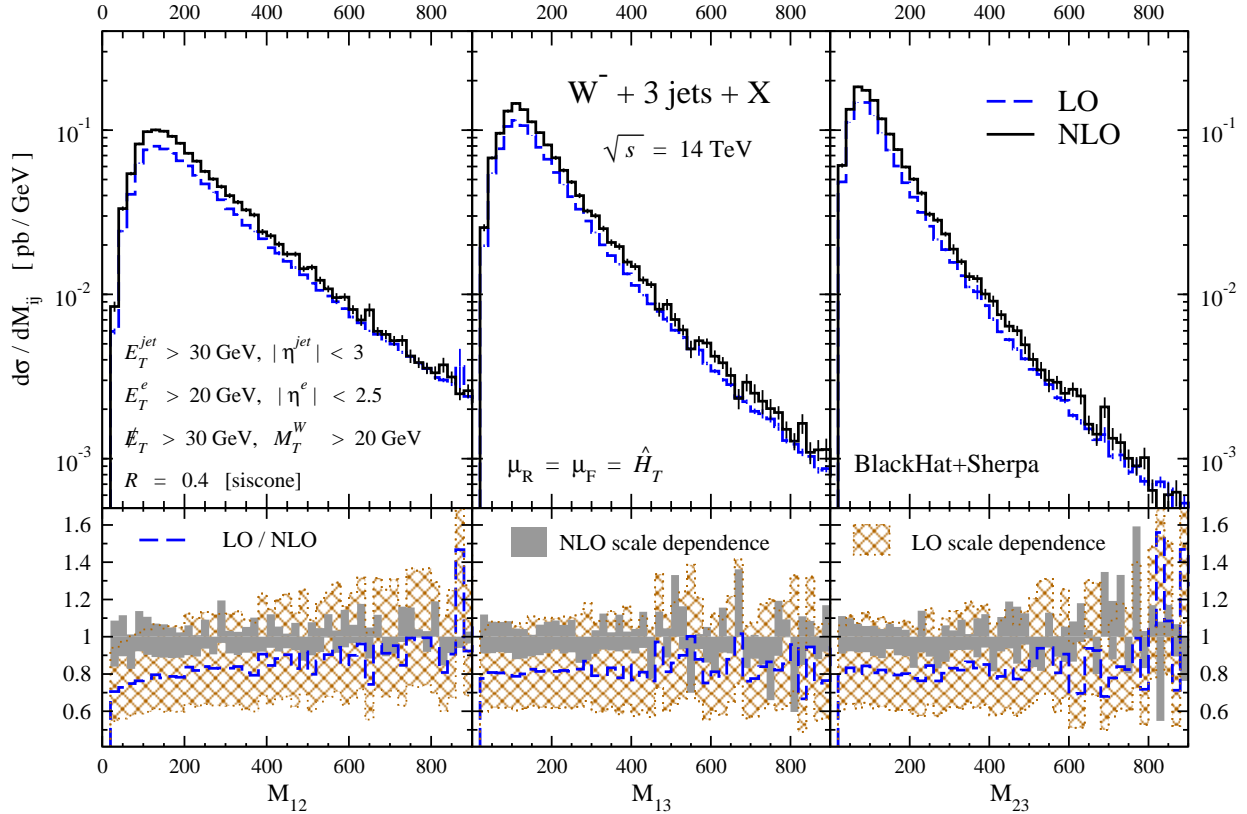}
\end{minipage}
\caption{The di-jet masses in $W^-+3$-jet production at the LHC; $M_{ij}$
(in GeV) is the invariant mass of the $i$-th and $j$-th leading jets, 
ordered in $E_T$.
}\label{DijetAllFigure}
\end{figure*}

\Fig{DijetAllFigure} displays distributions for the di-jet masses in
\Wmjjj-jet production.  The three plots in the figure give the di-jet
mass of the first and second, first and third, and second and third
leading jets, denoted by $M_{ij}$ where $i$ and $j$ label the jets.
Although our default choice of scale $\mu = \HTpartonic$ does
significantly reduce the shape changes between LO and NLO compared
with the choice $\mu = \ETW$ made for the Tevatron (see
\fig{W3j_DJMsFigure}), significant shape changes remain for the
$M_{12}$ distribution.  For the other two cases the ratio between LO
and NLO is rather flat.  These features have parallels in the $\Delta
R_{ij}$ distributions in \fig{Wm3jLHCHT_dRFigure}; the physics
of the two leading jets is not modeled especially well at LO.


\extraskip
\section{Leptons at the LHC}
\label{LeptonSection}

\begin{figure*}[tbh]
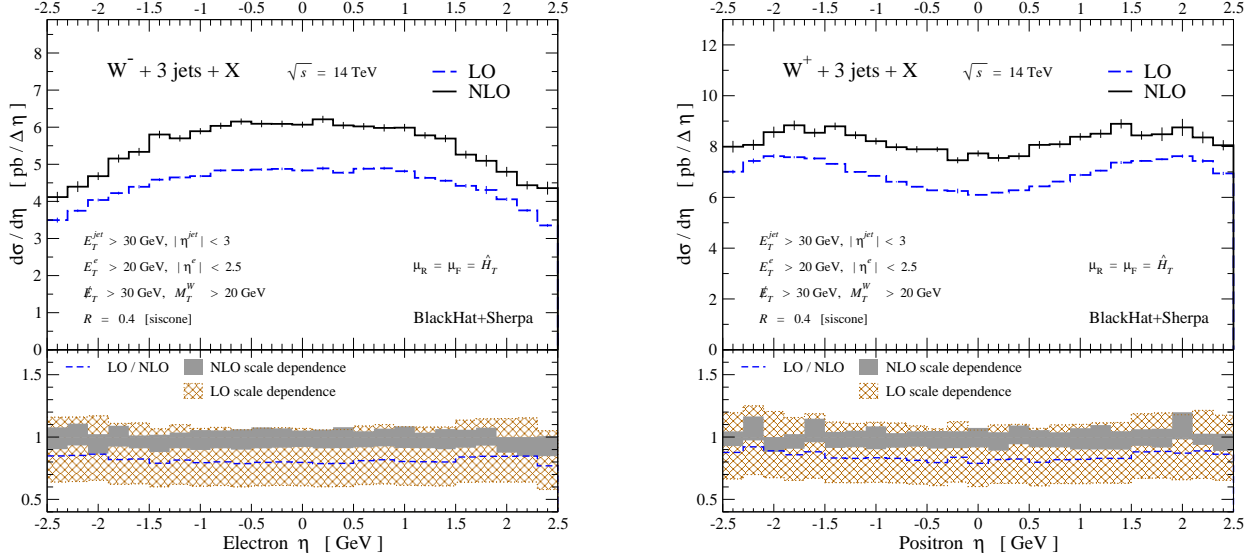

\begin{minipage}[b]{1.\linewidth}
\includegraphics[clip,scale=.37]{plots_final/Wm3jLHC_HTmu_siscone_eb_Etae-_A.eps}
\hfill
\includegraphics[clip,scale=.37]{plots_final/Wp3jLHC_HTmu_siscone_eb_Etae+_A.eps}
\end{minipage}
\caption{The charged-lepton pseudorapidity distribution at the LHC for
$W^-$ and $W^+$ production.
}
\label{WLeadingEtaLHCFigure}
\end{figure*}

We now turn from hadronic observables to leptonic ones.  At the LHC,
the latter distributions depend strongly on whether a $W^+$ or a $W^-$
boson has been produced.

\Fig{WLeadingEtaLHCFigure} shows the pseudorapidity distributions of
the daughter charged leptons.  Because of the large-$x$ excess of $u$
quarks over $d$ quarks, the $qg$ initial state produces $W^+$
preferentially, and tends to produce them more forward; this fact
accounts for the larger and more forward positron distribution.  The
lower panels show that in this case, the NLO corrections modify
primarily the overall normalization of these distributions, with only
a slight change in shape from LO to NLO.

\begin{figure*}[tbh]
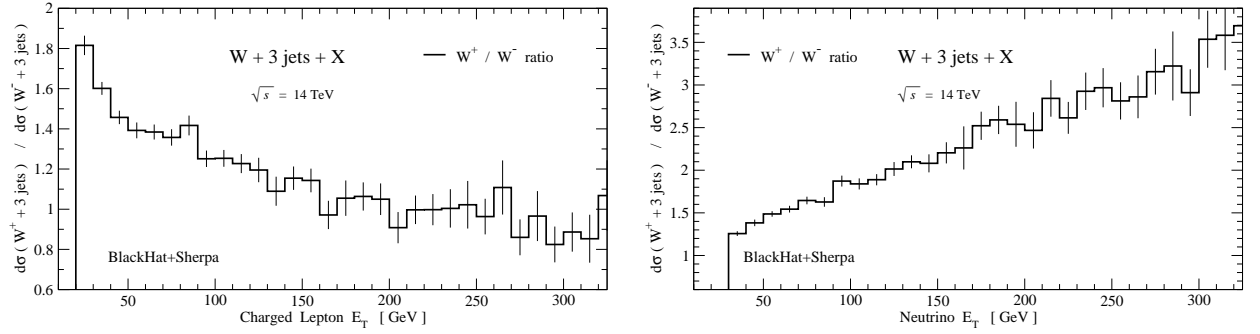

\begin{minipage}[b]{1.\linewidth}
\includegraphics[clip,scale=.34]{plots_final/Wp3jLHCHT_over_Wm3jLHCHT_central_mu_siscone_PTchargedlepton.eps}
\hfill
\includegraphics[clip,scale=.34]{plots_final/Wp3jLHCHT_over_Wm3jLHCHT_central_mu_siscone_PTneutrino.eps}
\end{minipage}
\caption{The left panel shows the ratio of the charged-lepton $E_T$
distributions at the LHC for $W^+$ and $W^-$ production, evaluated at
NLO, while the right panel shows the corresponding ratio for the
neutrino $E_T$, or equivalently the missing transverse energy.}
\label{WPTchargedneutralleptoni_ratio_Figure}
\end{figure*}

In the right panel of \fig{LeptonDistributions} we showed the positron
transverse momentum distribution in \Wpjjj-jet production at the LHC.
In order to contrast the distribution with the corresponding
distribution in \Wmjjj-jet production,
\fig{WPTchargedneutralleptoni_ratio_Figure} shows the ratio of the NLO
transverse energy distributions for the $W^\pm$ boson decay products,
charged leptons in the left panel and neutrinos in the right
panel. The plots show dramatic differences between the $W^+$ and $W^-$
distributions, especially for the neutrino $E_T$, or missing
transverse energy.  The left panel shows a large ratio for $W^+$ to
$W^-$ at small $E_T^e$ which declines at larger $E_T^e$.  In contrast,
the corresponding ratio for the neutrino $E_T$, or equivalently the
missing transverse energy $\ETslash$ in the event, starts with a
somewhat smaller value but increases rapidly with $E_T$.  This
significant increase means that the \Wpjn-jet background to
missing-energy-plus-jets signals, when a charged lepton is lost, is
more severe than the \Wmjn-jet background by a factor of two to three.

This disparate behavior is presumably due to a net left-handed
polarization for high $E_T$ $W^\pm$ bosons, which is then analyzed by
their leptonic decay via the parity-violating charged-current
interaction.  Before discussing this situation further, it is useful
to recall the dynamics underlying the {\it longitudinal} (rapidity)
charge asymmetry in $W^\pm$ production at the Tevatron, and the
corresponding asymmetry for the charged lepton into which the $W$
boson decays~\cite{ESW}.  At a $p\bar{p}$ collider, the dominance of
$u$ quarks over $\bar d$ quarks implies that in the process $u\bar{d}
\to W^+ \to e^+ \nu_e$ the $W^+$ typically moves in the $u$ quark
(proton) direction.  Because the charged current is left-handed, the
$u$ quark must be left-handed, and the $\bar{d}$ anti-quark
right-handed.  In order to conserve angular momentum, the $W^+$ must
be polarized left-handed along its direction of motion, which in this
case (at low transverse energy) is preferentially along the beam axis.
In the decay $W^+ \to e^+ \nu_e$, angular-momentum conservation
implies that the left-handed $W^+$ tends to emit the left-handed
neutrino forward, and the right-handed positron backward, relative to
its direction of motion.  The same arguments show that the $W^-$
typically moves in the anti-proton direction, is polarized
right-handed, and tends to decay with the left-handed electron
backward relative to its direction of motion.  Both signs of charged
leptons are typically more central than are their parent $W$ bosons.
In other words, there is a large asymmetry in the rapidity
distribution of $W^+$ bosons at the Tevatron, and a strongly diluted
asymmetry in the rapidity distribution of the charged decay
lepton~\cite{ESW,CDFWasym}.

Now consider a $pp$ collider (the LHC) and $W^\pm$ bosons moving with
large momentum primarily {\it transverse} to the beam axis, as
required to produce the large $E_T$ tails for the decay lepton
distributions shown in \fig{WPTchargedneutralleptoni_ratio_Figure}.
Suppose that both $W^+$ and $W^-$ bosons are polarized left-handed,
with a polarization that increases with $E_T^W$.  Then the $W^+$ will
tend to emit the left-handed neutrino forward relative to its
direction of motion (resulting in a larger transverse energy) and the
right-handed positron backward (smaller transverse energy).  In
contrast, the $W^-$ will emit the left-handed electron forward.  Such
decays will produce an enhancement in the neutrino $E_T$ distribution
and a depletion in the charged lepton distribution, for $W^+$ relative
to $W^-$, consistent with the ratios displayed
in~\fig{WPTchargedneutralleptoni_ratio_Figure}.  We have checked that
the same distributions as shown in
\fig{WPTchargedneutralleptoni_ratio_Figure}, but for \Wjx-jet
production, are very similar.  Also, the LO ratios in all cases are
virtually indistinguishable from the NLO ones.  We examined the LO
ratios, removing all lepton acceptance cuts, and the same general
trends persist (in fact they are even stronger at moderate lepton
$E_T$).  The left-handed polarization of both $W^+$ and $W^-$ is also
indicated by the corresponding ratios of $W$ transverse momenta (not
shown).  These ratios grow monotonically with the $W$'s $p_T$, but at
a lower rate than the ratio for the neutrino $E_T$.  This growth
reflects the fact that the larger the $W$ transverse momentum we
require, the larger is the required parton momentum fraction $x$, and thus
the more $W^+$ is favored over $W^-$ by the stiffer $u$ quark
distribution.

We do not have a complete understanding of why the $W$ bosons
should be polarized left-handed at large transverse momentum, 
in a manner that is apparently fairly independent of the number
of recoiling jets.  For \Wj-jet
production at LO, it is possible to examine the relevant helicity
amplitudes and make such an argument, based on kinematics and 
on the dominance of the $qg$ initial state.  It is also the case
that when a very high transverse-momentum (left-handed) quark
splits collinearly to another quark and a $W$ boson, with the 
$W$ boson taking most of the momentum, the $W$ boson is 
predominantly left-handed.  However, these examples certainly 
do not exhaust all of the possible polarization mechanisms,
and a more thorough explanation would require further study.


\extraskip
\section{Jet-Emission Probability at Increased Pseudorapidity Separation}
\label{RapiditySection}

One of the production mechanisms for the Higgs boson at the LHC
is via vector-boson fusion~\cite{VBF}, which contains partonic
subprocesses such as $qQ \to q' Q' H$, mediated by the fusion
of two $W$ bosons.  Because the Higgs is produced via colorless
electroweak vector boson exchange, a relative absence of radiation
is expected between the two forward tagging quark jets, in comparison
with QCD background processes with color exchange.
A veto on central jets may play an important role in such searches.
For this veto to be effective, the background processes should retain
a substantial probability of additional radiation, as the
pseudorapidity separation $\Delta\eta$
between two of the jets becomes large.
The production of a $W$ boson in association with jets is a prime example
of a background-type process dominated by color exchange at LO.
In fig.~61 of ref.~\cite{CHS},
a similar question was studied in \Wjj-jet and \Wjjj-jet
production at the Tevatron, by looking at the probability of finding
a third jet in the acceptance as a function of the pseudorapidity 
separation of the leading two jets, ordered by transverse energy. 
In that figure, CDF data
was compared with a leading-order QCD prediction.

\begin{figure*}[tbh]
\begin{minipage}[b]{1.\linewidth}
  \includegraphics[clip,scale=0.55]{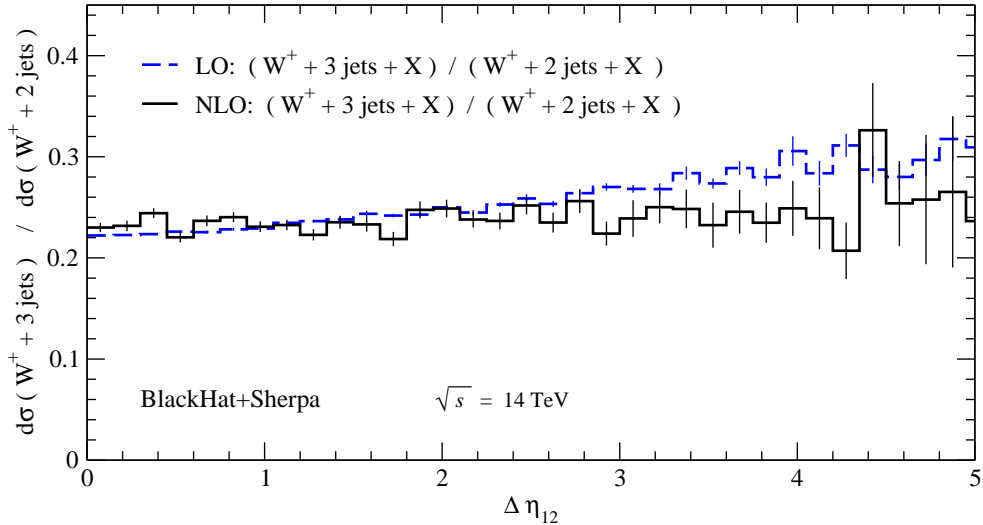}
\end{minipage}
\caption{The ratio of the inclusive \Wpjjj-jet to \Wpjj-jet cross
sections at NLO, as a function of the pseudorapidity
separation $\Delta\eta_{12}$
between the leading two jets at the LHC.  The solid (black)
line gives the NLO result, while the dashed (blue) line represents the
LO result.  The ratio for $W^-$ is very similar, particularly at NLO.}
\label{DiJetDETaRatioFigure}
\end{figure*}

A more appropriate distribution for assessing the effectiveness of a
central jet veto would be to order the jets in pseudorapidity, not
$E_T$, and place an additional constraint that the third jet be
between the two most widely-separated jets in pseudorapidity, but here
we match the choice made in ref.~\cite{CHS}.  
In \fig{DiJetDETaRatioFigure} we plot the ratio of the
\Wpjjj-jet to \Wpjj-jet cross sections at the LHC, as a function of
the pseudorapidity separation of the leading two jets, ordered in
$E_T$.  This ratio measures the emission probability of a third jet.
The solid line gives the NLO equivalent of the LO Tevatron results in
fig.~61 of ref.~\cite{CHS}.  As was found at the Tevatron, the
emission probability is substantial, over 20\%, and remarkably
independent of $\Delta\eta_{12}$.  Although we plot only the emission
probability for jets accompanying a $W^+$, the corresponding plot for
$W^-$ is essentially indistinguishable from it at NLO.  (The
difference between the LO and NLO results for $W^-$ is smaller
than the difference shown in~\fig{DiJetDETaRatioFigure} for $W^+$.)


\extraskip
\section{Subleading-Color Terms}
\label{ColorApproximationSection}

In this section, we turn our attention to the question of simplifying
a computation by taking advantage of the structure of color sums.  As
explained in \sect{DetailsSection}, we can organize the matrix
elements---leading-order, real-emission, or virtual---in an expansion
in $1/N_c$.  We expect higher-order terms in this expansion to give
smaller contributions numerically; but there are more of them, and
their structure is more intricate than that of lower-order terms. In
general they take significantly more computer time per event to
evaluate.  Although one could simply drop these contributions once
they have been shown---preferably by direct computation---to be
negligible or reliably estimated, we shall describe how to reduce the
computer time their direct computation would entail.

In ref.~\cite{PRLW3BH}, we used a particular type of ``leading-color''
approximation (LC NLO), in which a subset of subleading-color terms were
dropped.  In the real-emission contributions, as well as in the
real-subtraction terms, we retained all terms in the color
expansion; the same was also true for the singular terms in the
virtual matrix elements.  The only approximation was within the finite
virtual terms.  Here, by ``finite'' we mean the $\e^0$ term in the
Laurent expansion of the infrared-divergent one-loop amplitudes in $\e
= (4-D)/2$, after extracting a multiplicative factor of
$c_\Gamma(\e)$ (defined in \eqn{cGammaDef}).

The approximation was defined by first dropping the subleading-color
terms, that is those suppressed either by powers of $1/N_c^2$
(including those coming from leading-color partial amplitudes) or
$n_f/N_c$ (the latter arising from virtual quark loops), in the ratio
of the finite virtual terms to the tree-level cross section.  In a
second step, we multiply this truncated ratio by the tree cross
section, with its full color dependence.  The net effect of this
approximation is to drop quark loops and subleading-color terms in the
finite virtual terms that have a different kinematic structure than
those at tree level, while retaining the subleading-color terms that
have the same kinematic structure.  This approximation turns out to be
a much better estimate than a strict leading-color approximation
(dropping all subleading-color terms in the real-emission terms as
well), while still simplifying the calculation considerably by
eliminating the need to compute primitive one-loop amplitudes that
contribute only to subleading-color terms, such as those shown in
\fig{LoopDiagramsSLCFigure}.  

\begin{table}
\vskip .4 cm
\begin{tabular}{|c||c|c|c|}
\hline
number of jets  & CDF &  LC NLO & NLO  \\
\hline
1  & $\; 53.5 \pm 5.6 \;$ & $\; 58.3(0.1)^{+4.6}_{-4.6} \;$ &
            $\;  57.83(0.12)^{+4.36}_{-4.00}  \;$ \\
\hline
2  & $6.8 \pm 1.1$  & $ 7.81(0.04)^{+0.54}_{-0.91}$ &
            $7.62(0.04)^{+0.62}_{-0.86} $  \\
\hline
3 &  $0.84\pm 0.24$  & $\; 0.908(0.005)^{+0.044}_{-0.142} \;$
   & $0.882(0.005)^{+0.057}_{-0.138}$  \\
\hline
\end{tabular}
\caption{Comparison of LC NLO to full NLO for the total inclusive cross
sections in pb of \Wjn-jet production at the Tevatron using CDF's
cuts~\cite{WCDF} ($E_T^{n\rm th\hbox{-}jet} > 25$ GeV) and the
\SISCone{} algorithm.  For reference (see also
\tab{CDFCrossSectionTable}), the first column gives the CDF data.  The
second column shows the LC NLO results and the third column the
complete NLO results.
\label{CDFLCNLOCrossSectionTable} }
\end{table}

\Tab{CDFLCNLOCrossSectionTable} compares NLO results for the total
cross sections at the Tevatron with the experimental setup as in
\eqns{TeVJetCuts}{TeVLeptonCuts} except for the tighter jet cut,
$E_T^\jet > 25\,{\rm GeV}$.  For reference, we also show the
corresponding CDF data.  The column labeled ``LC NLO'' contains the
results computed using the specific leading-color approximation of
ref.~\cite{PRLW3BH}.  The last column gives the full NLO result,
incorporating all subleading-color terms.  Previously~\cite{PRLW3BH},
we showed explicitly that this approximation is very good for \Wjx-jet
production at the Tevatron, leading to errors of no more than three
percent.  The entry for \Wjjj{} jets is new, and demonstrates that
just as for \Wj-jet and \Wjj-jet production, the LC approximation is
excellent, shifting the total cross section by just under three
percent.  This shift is much smaller than the NLO scale dependence.
In all cases, the LC NLO and complete NLO result are both in excellent
agreement with the data.

\begin{table}
\vskip .4 cm
\begin{tabular}{|c||c|c||c|c|}
\hline
cut  & $\; W^-$ LC NLO$\;$ & $\;W^-$ NLO$\;$ & $\; W^+$ LC NLO$\;$
    & $\; W^+$ NLO$\;$ \\
\hline
$\;E_T^\jet>30$ GeV$\;$
  & $\; 28.17(0.13)^{+0.99}_{-2.18} \;$
  & $\; 27.52(0.14)^{+1.34}_{-2.81} \; $
  & $\; 42.33(0.27)^{+1.82}_{-2.68} \;$
  & $\; 41.47(0.27)^{+2.81}_{-3.50}$ \\
\hline
$\; E_T^\jet>40$ GeV$\;$
  &  $14.24(0.07)^{+0.76}_{-1.09}$
  &  $13.96(0.07)^{+1.03}_{-1.31}$
  &  $22.08(0.15)^{+1.20}_{-1.44}$
  &  $21.76(0.15)^{+1.68}_{-1.86}$  \\
\hline
\end{tabular}
\caption{Comparison of the total cross sections, in pb, between LC NLO
and full NLO results for \Wjjj-jet prodiction at the LHC with
$\sqrt{s} = 14$~TeV, using the \SISCone{} algorithm and
the cuts of \eqns{LHCCuts}{LHCJetCut}.  }
\label{LHCCrossSectionTable}
\end{table}

\Tab{LHCCrossSectionTable} shows results for the total cross section
of \Wjjj-jet production at an LHC energy of 14~TeV, using the cuts
given in \eqns{LHCCuts}{LHCJetCut}.  In this case the column
labeled ``LC NLO'' refers to an LC approximation that is slightly
modified from the one used for the Tevatron.  We avoid rescaling the
leading-color virtual contributions by the ratio of the full-color to
leading-color cross section; this allows us to simply add
together the ``leading-color'' and remaining ``subleading-color''
contributions to obtain the total cross section.  This modification
has only a small effect on the corresponding total cross section.
Indeed, for Tevatron cross sections the shift is under 1.5 percent
throughout the scale-variation band.  (To facilitate comparison to our
previously published results~\cite{PRLW3BH}, for Tevatron cross
sections we use the identical LC approximation as in that reference.)

\begin{figure*}[tbh]
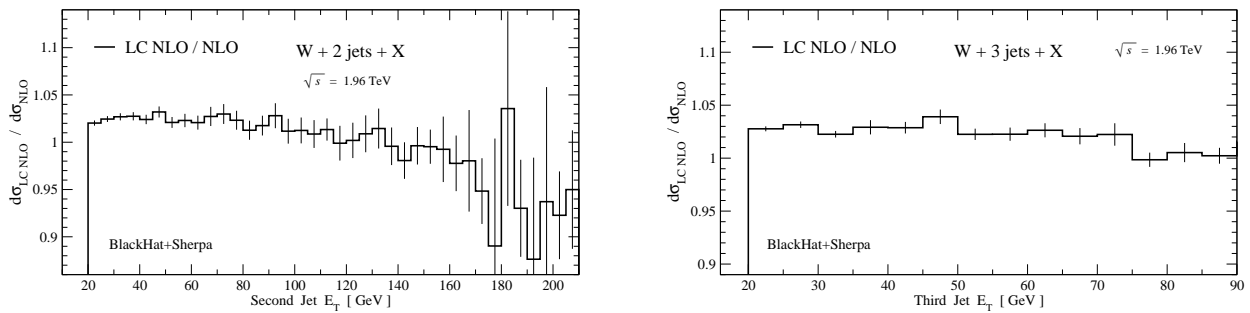

\begin{minipage}[b]{1.0\linewidth}
\includegraphics[clip,scale=0.32]{plots_final/W2jTev_ETWmu_siscone_lc_fc_comparison_eb_jets_jet_1_1_Et_2.eps}
\hfill
\includegraphics[clip,scale=0.32]{plots_final/W3jTev_ETWmu_siscone_lc_fc_comparison_eb_jets_jet_1_1_Et_3.eps}
\end{minipage}
\caption{Ratios of our leading-color approximation to a full-color
calculation for $E_T$ distributions in \Wjjx-jet production at the
Tevatron.  The numerical integration errors on each bin are indicated 
by thin vertical lines. (Different error sources are combined linearly.)}
\label{LCNLOTevRatiosFigure}
\end{figure*}

The modified LC approximation is again accurate to three
percent for central values, and to five percent for the upper or lower
edges of the scale-dependence bands.  The good quality of the LC
approximation also holds for all distributions we have examined.
Examples are shown in \figs{LCNLOTevRatiosFigure}{LCNLOLHCRatiosFigure};
for both the Tevatron and the LHC, the corrections due to the subleading-color
terms are less than three percent, uniformly across the distributions.

\begin{figure*}[tbh]
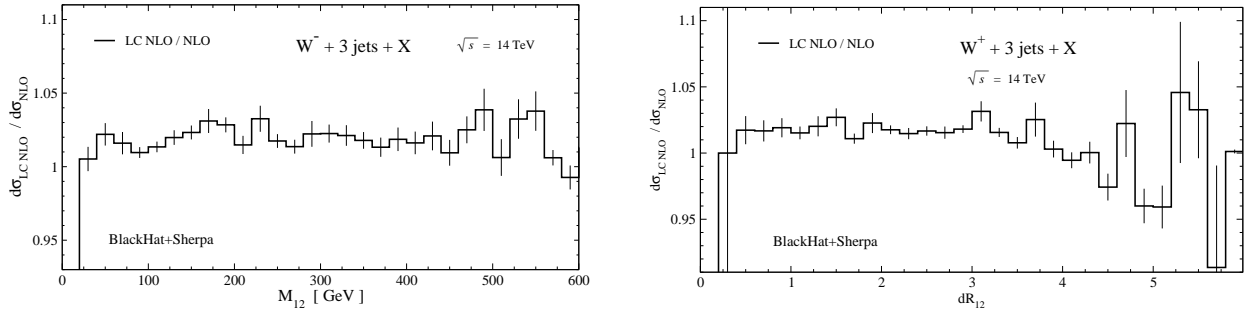

\begin{minipage}[b]{1.\linewidth}
\includegraphics[clip,scale=0.32]{plots_final/Wm3jLHC_HTmu_siscone_lc_fc_comparison_eb_jets_jet_1_1_DJM_B1.eps}
\hfill
\includegraphics[clip,scale=0.32]{plots_final/Wp3jLHC_HTmu_siscone_lc_fc_comparison_eb_jets_jet_1_1_dR2__A1.eps}
\end{minipage}
\caption{Ratios of our leading-color approximation to a full-color
calculation for the leading-dijet invariant mass
distribution in \Wmjjj-jet production,
and for the $\Delta R_{12}$ distribution in \Wpjjj-jet
production, at the LHC.}
\label{LCNLOLHCRatiosFigure}
\end{figure*}

The quality of this approximation has important implications for
organizing the calculation of virtual contributions.  As discussed in
\sect{ss_ColorOrganization}, the primitive amplitudes entering the
leading-color approximation used in ref.~\cite{PRLW3BH} and discussed
above are a small subset of the primitive amplitudes required for the
complete virtual correction.  The full result, including
subleading-color terms, requires the computation of a much larger
number of primitive amplitudes, 336 for two-quark processes and 80 for
four-quark processes\footnote{The count for the four-quark case
is for non-identical fermions;
for identical fermions, multiply by 2.} as opposed to 48 for two-quark
processes and 8 for four-quark processes in the leading-color
computation.  Furthermore, the subleading-color primitive amplitudes
shown in \fig{LoopDiagramsSLCFigure} have a more complicated analytic
structure than the leading-color ones shown in
\fig{LoopDiagramsLCFigure},  because the $W$ does not have to
be ordered with respect to parton(s) emitted between the $W$ boson and
the $q$ or $q'$.  The partial lack of color-ordering implies cuts
and poles in more channels.  As a result, it takes about 27
times longer per phase-space point to evaluate the subleading-color
virtual terms than the leading-color ones.

This large factor may seem to be a cause for concern.  However, the
smallness of the subleading-color contributions, discussed above,
comes to our rescue.  If we require that the numerical integration
errors due to the subleading-color contributions be comparable to
those coming from the leading-color ones, we can allow for larger {\it
relative\/} errors in the evaluation of the subleading-color terms.
We can adopt a ``color-expansion sampling'' approach wherein we can
use far fewer phase-space points (typically a factor of 20 fewer) to
evaluate them~\cite{NLOZ4Jets} as compared to the leading-color terms.
(We must ensure that there are sufficient statistics in each bin of
every distribution of interest, of course.)  There is no need to know
ahead of time what the relative size of the two contributions is; we
simply stop the integration when the desired numerical precision is
reached for each contribution separately.  This approach requires only
a bit more than a factor of two more computer time for the full-color result
than for the leading-color approximation, with our present setup.  It
saves roughly a factor of thirty in computer time, compared to the
naive approach of evaluating the subleading-color terms at every
phase-space point.  We expect to obtain further improvements in the
evaluation efficiency of subleading-color contributions through
improved re-use of primitive amplitudes.  Together with
color-expansion sampling, this should reduce the time required for
computing the subleading-color terms to a small fraction of the total
computer time.  This color-expansion sampling approach would naturally
be implemented in a dynamical way by treating the subleading-color
contributions as another set of `subprocesses' within the \SHERPA{}
multi-channel integration.

Were we seeking to optimize the computer time more aggressively,
we would let the total error be {\it dominated\/} by the most
time-consuming part of the calculation, namely the subleading-color
terms.  We have, however, opted for a more conservative approach,
keeping the integration error from the subleading-color contributions
in line with the errors from the other contributions.


\extraskip
\section{Conclusions}
\label{ConclusionSection}

In this paper, we presented the first complete NLO study of \Wjjj-jet
production, incorporating all massless partonic contributions.  We
compared the total cross section and the third-jet $E_T$ distribution to
Tevatron data~\cite{WCDF}. We also presented NLO predictions for the $H_T$
and di-jet mass distributions.  It will be interesting to compare these
and other distributions to Tevatron data sets with larger statistics than
those already published by CDF~\cite{WCDF}.  We
presented a variety of distributions at the declared final running
energy of the LHC, including many relevant for Standard Model
backgrounds to events with large missing energy and
to Higgs boson production via vector-boson fusion.
As expected, we find a much smaller
renormalization- and factorization-scale dependence in all
distributions at NLO, compared to LO results.  Although the LHC will
start running at lower energy, our choice should help facilitate
comparisons to earlier studies based on leading-order QCD and
matching to parton showers~\cite{LOPrograms, HELAC, Amegic,
Matching,MLMSMPR,OtherLOWCodesPartonShowers,LOComparison}.

We have shown explicitly in \Wjjj{} jet distributions that the scale
dependence of LO predictions is not restricted to overall
normalizations.  An infelicitous choice of scale can change the shapes
of distributions substantially between LO and NLO.  This effect is
much more pronounced at the LHC than at the Tevatron.  One can reduce
the change in shape of distributions between LO and NLO by choosing a
scale dynamically, event by event, corresponding to a typical scale
for the event, as noted in, for example,
refs.~\cite{Vplus1NLOAR,EarlyWplus2MP,DynamicalScaleChoice}.  The
problem with poor scale choices can be much more severe than just
changes in shape between LO and NLO results.  Indeed, for sufficiently
poor choices, such as the fixed scale $\mu = M_W$ or the transverse
energy of the $W$ boson, $\mu = E_T^W$, large logarithms can appear
in some distributions, invalidating even an NLO prediction.
We find that the total (partonic) transverse energy $\HTpartonic$ is a
more appropriate scale choice for \Wjjj-jet production than the $W$
transverse energy or the fixed scales used in previous Tevatron
analyses.  (A fixed fraction of the total transverse energy would also
be appropriate.)  We expect that this scale choice will be appropriate
to a variety of higher-multiplicity processes, and recommend its use
in LO predictions (when an NLO one is not available) as well as at
NLO.  A recent paper~\cite{Bauer} motivates a similar type of scale
choice using soft-collinear effective theory, and we have contrasted
its properties with those of $\HTpartonic$.  Of course, a simple scale
choice is no substitute for a complete NLO prediction.  In some
distributions, such as the transverse energy of the second-most
energetic jet and the $\Delta R$ separation between the two leading
jets, the NLO calculation incorporates physics effects that are not
captured by simple changes of scale.

We also confirmed that our previous NLO analysis of \Wjjj-jet
production~\cite{PRLW3BH}, which used a specific leading-color
approximation, is valid to within three percent.  This error is quite
a bit smaller than other uncertainties, such as that implicit in the
scale dependence, or that due to uncertainties in the parton
distribution functions.  However, we can draw this conclusion only
after computing the subleading-color terms, as we have done here.  To
evaluate the subleading-color terms efficiently, we used
``color-expansion sampling''.  The subleading-color terms require much
more computer time per phase space point.  However, because they are
small, only a few percent of the leading-color ones, we can tolerate a
much larger relative error for them from the Monte Carlo integration,
thus sampling them much less often.  We expect that this general
approach will be an effective technique for reducing the computer-time
requirements for ever-more complicated processes such as \Zjjj-jet,
\Wjjjj-jet, or \Zjjjj-jet production.

In our analyses we mainly used the \SISCone{} jet algorithm; we also
presented total cross sections using the $k_T$ jet
algorithm~\cite{KTAlgorithm} at the LHC.  These jet algorithms are
infrared-safe to all orders in perturbation theory.  With our setup it
is a simple matter to replace one infrared-safe cone algorithm with
any other desired one.  We defer a study of the anti-$k_T$
algorithm~\cite{antikT}, which has certain experimental advantages
such as uniform catchment areas for soft radiation, to future work.
From a perturbative viewpoint, infrared safety is essential;
infrared-unsafe quantities are simply logarithmically divergent.  In
the real world, perturbation theory does not go on forever but is
overtaken by non-perturbative dynamics around the confinement scale.
Infrared-unsafe quantities are not infinite, but the infinities are
cut off and replaced by quantities determined by non-perturbative
physics.  The logarithms translate~\cite{SISCONE,GavinSalamThesis}
into inverse powers of the strong coupling $\alpha_S$, thereby
spoiling the perturbative expansion.  This is an important practical
problem because the jet algorithms traditionally used at the Tevatron
by the CDF and D0 collaborations are, in fact, infrared-unsafe beyond
the lowest orders~\cite{SISCONE}.  Unknown nonperturbative corrections
for these algorithms would undo many of the benefits of a higher-order
prediction, especially in the context of new, higher statistics data.
Accordingly, it is highly desirable that future experimental analyses
at both the Tevatron and the LHC use an infrared-safe jet algorithm.

Our paper demonstrates the utility of on-shell methods
for computing one-loop matrix elements entering state-of-the-art
NLO QCD predictions for processes of phenomenological interest
at the LHC.  We used \BlackHat{}, an efficient new
code library based on these methods. The NLO \Wjjj-jet results
reported here also demonstrate the functionality of our computational setup,
which uses
\BlackHat{} in conjunction with the \SHERPA{} package.  Besides
handling the real-emission contributions and infrared-singular phase
space via \AMEGIC{}, the \SHERPA{} framework offers a convenient set
of tools for integrating over phase space and analyzing the results.

There are many relevant processes with large numbers of final-state
objects such as jets that remain to be computed, especially those
involving vector bosons, jets, heavy quarks and Higgs
bosons~\cite{LesHouches}.  Such processes are backgrounds to the
production of new heavy particles with multi-body decays.
Our setup is robust enough to deal systematically with such processes.
In the present paper, we have demonstrated the new
tools and on-shell methods at work for the non-trivial case of
\Wjjj-jet production at hadron colliders.  We look forward to
comparing our predictions against forthcoming LHC data.

\extraskip
\section*{Acknowledgments}

We thank Jeppe Andersen,
Christian Bauer, John Campbell, Keith Ellis, Beate Heinemann,
Joey Huston, Pavel Nadolsky, Michael Peskin, Gavin Salam, Rainer
Wallny and Giulia Zanderighi for helpful discussions.  We especially
thank Costas Papadopoulos and Roberto Pittau for assistance in
comparing results for the virtual contributions to squared matrix
elements.  This
research was supported by the US Department of Energy under contracts
DE--FG03--91ER40662, DE--AC02--76SF00515 and DE--FC02--94ER40818.
DAK's research is supported by the European Research Council under
Advanced Investigator Grant ERC--AdG--228301.  
HI's work is supported by a fellowship from the US LHC Theory
Initiative through NSF grant PHY-0705682.  This research used
resources of Academic Technology Services at UCLA, PhenoGrid using the
GridPP infrastructure, and the National Energy Research Scientific
Computing Center, which is supported by the Office of Science of the
U.S. Department of Energy under Contract No. DE-AC02-05CH11231.

\extraskip
\appendix

\section{Squared Matrix Elements at One Point in Phase Space}
\label{MatrixElementAppendix}

In order to aid future implementations of virtual corrections for
\Wjjj-jet production in other numerical codes, we present values of
the one-loop virtual corrections to the squared matrix elements,
$\MNLO$, at one point in phase space.  This comes from 
the interference between the tree and one-loop
amplitudes, summed over all colors and helicities, for $N_c=3$ and
$n_f=5$ massless quark flavors.

In \tab{ME2Table} we present numerical values for four representative
subprocesses.  All other subprocesses are related to these four by
crossing symmetry.  In the second and third lines of \tab{ME2Table},
the presence of two identical quarks (after crossing all particles
into the final state) means that amplitudes are antisymmetrized
under exchange of the two.

\begin{table}[htp]
\vskip .4 cm
\begin{tabular}{|c||c|c|r|}
\hline
$\hatMNLO$ &$1/\epsilon^2$&$1/\epsilon$ &
 $\epsilon^0 \hskip .95 cm $ \\\hline\hline
$(1_{\bar u}\, 2_{c} \rightarrow 3_c\, 4_{\bar d}\, 5_g\, 6_{e^-} 7_{\nub})$ &
$ -8.333333333$&
$-32.37677210$&
$  1.778061330$\\\hline
$(1_{\bar u}\, 2_u \rightarrow 3_u\, 4_{\bar d}\, 5_g\, 6_{e^-} 7_{\nub})$&
$ -8.333333333$&
$-32.40807165$&
$  1.035000256$\\\hline
$(1_{\bar u}\, 2_d \rightarrow 3_d\, 4_{\bar d}\, 5_g\, 6_{e^-} 7_{\nub})$&
$ -8.333333333$&
$-32.50750136$&
$  0.4788030624$\\\hline
$(1_{\bar u}\, 2_g \rightarrow 3_g\, 4_g\, 5_{\bar d}\, 6_{e^-} 7_{\nub})$&
$-11.66666667$&
$-42.34303628$&
$-13.97991225$\\\hline
\end{tabular}
\caption{
Numerical values of the normalized virtual correction to
the squared matrix elements,
$\hatMNLO$, at the phase-space point
given in the text, for the four basic
partonic subprocesses for \Wjjj-jet production at a hadron collider.
We give the finite parts along with the coefficients
of the poles in $\epsilon$.
}
\label{ME2Table}
\end{table}

We quote numerical results for the ultraviolet-renormalized virtual
corrections in the 't~Hooft-Veltman variant of dimensional
regularization~\cite{HV}.  The remaining singularities in the
dimensional regularization parameter $\epsilon = (4-D)/2$ arise from
the virtual soft and collinear singularities in the one-loop
amplitudes.

The quoted values are for the ratio of the virtual corrections to the
tree-level squared matrix element $\MLO$.  Explicitly, we define the
ratio,
\begin{equation}
  \hatMNLO
\equiv
\frac{1}{8\pi\alpha_S \, c_\Gamma(\e)}
\frac{\MNLO}{\MLO}\,,
  \label{ME2normalization}
\end{equation}
where we have also separated out the dependence on the
strong coupling $\alpha_S$ and the overall
factor $c_\Gamma(\e)$, defined by
\begin{equation}
c_\Gamma(\e)
= \frac{1}{(4\pi)^{2-\epsilon}}
\frac{\Gamma(1+\epsilon)\Gamma^2(1-\epsilon)}
 {\Gamma(1-2\epsilon)}\,.
  \label{cGammaDef}
\end{equation}

The coupling constants, mass and width of the $W$ boson are given in
\sect{CouplingsSection}.  However, the numerical values for the
ratio~(\ref{ME2normalization}), given in \tab{ME2Table}, are
independent of these parameters; coupling constants as well as the $W$
boson Breit-Wigner factor cancel between the tree and 
virtual correction terms.

We choose the phase-space point given in eqs. (9.3) and (9.4) of
ref.~\cite{Genhel},
\begin{eqnarray*}
k_1 &=&  {\mu \over 2}\, (1, -\sin \theta,
           -\cos \theta \sin \phi, -\cos \theta \cos\phi)\,, \\
k_2 &=&  {\mu \over 2}\, (1,  \sin \theta,
          \cos \theta \sin \phi,  \cos \theta \cos \phi)\,,  \\
k_3 &=&  {\mu\over 3} (1,1,0,0) \,, \\
k_4 &=&  {\mu\over 8} (1, \cos\beta, \sin \beta,0) \,,  \\
k_5 &=&  {\mu \over 10} (1, \cos\alpha \cos\beta, \cos \alpha \sin \beta,
                                 \sin \alpha )\,,  \\
k_6 &=& {\mu \over 12} (1, \cos\gamma \cos\beta,
            \cos \gamma \sin \beta, \sin \gamma)\,, \\
k_7 &=& k_1+k_2-k_3-k_4-k_5-k_6\,,
\label{SevenPointKinematics}
\end{eqnarray*}
where
\begin{eqnarray*}
 \theta = {\pi\over 4}\,,\hskip 1 cm
 \phi   =  {\pi\over 6}\,,\hskip 1 cm
 \alpha = {\pi \over 3}\,,\hskip 1 cm
 \gamma = {2 \pi \over 3} \,, \hskip 1 cm
 \cos \beta = - {37\over 128} \,,
\end{eqnarray*}
and the renormalization scale $\mu$ is set to $\mu = 7$~GeV.  We have
flipped the signs of $k_1$ and $k_2$ compared to
ref.~\cite{Genhel}, to correspond to $2\rightarrow 5$
kinematics, instead of $0\rightarrow 7$ kinematics.
The labeling of the parton and lepton momenta is indicated
explicitly in the first column of Table~\ref{ME2Table}.


\end{document}